\documentclass[twocolumn,superscriptaddress,numerical,showpacs,amsmath,amssymb,aps,pra,floatfix]{revtex4-1}

\usepackage{bbm}
\usepackage{amsfonts}

\usepackage{blindtext}

\usepackage{amsmath}    %
\usepackage{braket}
\usepackage{graphicx}   %
\usepackage{verbatim}   %
\usepackage{color}      %
\usepackage{subfigure}  %
\usepackage{hyperref}   %
\newcommand\norm[1]{\left\lVert#1\right\rVert}

\renewcommand{\Re}{\operatorname{Re}}
\renewcommand{\Im}{\operatorname{Im}}

\begin{document}

\title{Error-correction and noise-decoherence thresholds for coherent errors in planar-graph surface codes}

\author{F. Venn}
\affiliation{DAMTP, University of Cambridge, Wilberforce Road, Cambridge, CB3 0WA, United Kingdom}
\author{B. B\'eri}
\affiliation{DAMTP, University of Cambridge, Wilberforce Road, Cambridge, CB3 0WA, United Kingdom}
\affiliation{T.C.M. Group, Cavendish Laboratory, University of Cambridge, J.J. Thomson Avenue, Cambridge, CB3 0HE, United Kingdom}

\begin{abstract}
We numerically study coherent errors in surface codes on planar graphs, focusing on noise of the form of $Z$- or $X$-rotations of individual qubits. 
We find that, similarly to the case of incoherent bit- and phase-flips, a trade-off between resilience against coherent $X$- and $Z$-rotations can be made via the connectivity of the graph. 
However, our results indicate that, unlike in the incoherent case, the error-correction thresholds for the various graphs do not approach a universal bound. 
We also study the distribution of final states after error correction. 
We show that graphs fall into three distinct classes, each resulting in qualitatively distinct final-state distributions. In particular, we show that a graph class exists where the logical-level noise exhibits a decoherence threshold slightly above the error-correction threshold. In these classes, therefore, the logical level noise above the error-correction threshold can retain significant amount of coherence even for large-distance codes. To perform our analysis, we develop a Majorana-fermion representation of planar-graph surface codes and describe the characterization of logical-state storage using fermion-linear-optics-based simulations. 
We thereby generalize the approach introduced for the square lattice by Bravyi \textit{et al}. [npj Quantum Inf. 4, 55 (2018)] to surface codes on general planar graphs.
\end{abstract}

\maketitle

\section{Introduction}
In recent years, significant progress has been made to improve the coherence times of qubits \cite{barendsSuperconductingQuantumCircuits2014a, yanFluxQubitRevisited2016, kjaergaardSuperconductingQubitsCurrent2020}, including demonstrations of key ingredients for quantum error correction (QEC)~\cite{calderbankGoodQuantumErrorcorrecting1996, steaneErrorCorrectingCodes1996,kellyStatePreservationRepetitive2015, takitaDemonstrationWeightFourParity2016, terhalQuantumErrorCorrection2015}.
To proceed further on the way to long-time stable qubits, topological codes such as the surface code \cite{bravyiQuantumCodesLattice1998,kitaevFaulttolerantQuantumComputation2003, fowlerSurfaceCodesPractical2012} are considered promising candidates.

One of the major benefits of the surface code is its high tolerance to errors in the physical qubits~\cite{dennisTopologicalQuantumMemory2002, fowlerSurfaceCodesPractical2012}. Error rates at the theoretically estimated fault-tolerance threshold have already been reached in experiments~\cite{barendsSuperconductingQuantumCircuits2014a}. These thresholds are usually based on the assumption that the noise acts in the form of Pauli noise, an error model in which the action on the physical qubits is given by Pauli operators chosen from a probability distribution. In the uncorrelated case, the action on single qubits can be described by the channel 
\begin{equation}
	\mathcal{E^\text{P}[\rho]} = (1 - \epsilon) \rho + \epsilon_x X \rho X + \epsilon_y Y \rho Y + \epsilon_z Z \rho Z,
\end{equation}
where $\rho$ is the state of the qubit, $X, Y, Z$ denote the Pauli operators, and $ \epsilon_x, \epsilon_y, \epsilon_z$ are suitably chosen probabilities ($\epsilon=\sum_j\epsilon_j$).
This channel is also referred to as incoherent single-qubit error. Based on this error model, analytic results show that an error threshold exists under which, by increasing the number of qubits in the code, the error rate for qubits encoded in the code (so-called logical qubits) can be made arbitrary small~\cite{dennisTopologicalQuantumMemory2002}.
The appeal of the incoherent error model is that all operations are from the Clifford group.
This, together with the stabilizer code nature of the surface code, implies that the effect of such errors can be efficiently simulated classically according to the Gottesmann-Knill theorem~\cite{aaronsonImprovedSimulationStabilizer2004}. This allowed the numerical establishment of high threshold rates~\cite{fowlerHighthresholdUniversalQuantum2009, wangSurfaceCodeQuantum2011}, which gives reason for optimism that QEC and ultimately general quantum computation is achievable.

One of the limitations of the Pauli error model is that it does not include ``coherent noise'', e.g., errors where each qubit undergoes a unitary rotation. 
These kinds of errors inevitably occur (e.g., due to qubit detuning) in  quantum devices and therefore their interplay with QEC procedures needs to be understood. 
Mathematically, focusing on single-qubit errors, coherent errors correspond to the error channel
\begin{equation}
	\mathcal{E^\text{c}[\rho]} = U \rho U^\dagger,
\end{equation}
with $U \in SU(2)$.

Theoretical studies of coherent errors suggest that they act substantially differently from incoherent errors~\cite{aliferisQuantumAccuracyThreshold2006, chamberlandHardDecodingAlgorithm2017, gutierrezErrorsPseudothresholdsIncoherent2016, caiMitigatingCoherentNoise2020}.
In certain circumstances, they can build up quadratically faster than incoherent errors~\cite{gottesmanMaximallySensitiveSets2019}.
It has been shown that they affect average fidelities less than incoherent errors, but introduce higher diamond-norm error rates~\cite{sandersBoundingQuantumGate2015}. 
On the other hand, the logical-level diamond-norm error rate can scale with code distance as a more favorable power of the physical-qubit diamond-norm error for coherent than for incoherent errors~\cite{huangPerformanceQuantumError2019}. 
It was also found that even if physical qubits experience coherent errors, the logical-level noise, especially upon averaging over error-syndromes, becomes increasingly incoherent with increasing code distance~\cite{greenbaumModelingCoherentErrors2018,bravyiCorrectingCoherentErrors2018,bealeQuantumErrorCorrection2018}, 
however quantifying this has some subtleties~\cite{iversonCoherenceLogicalQuantum2020}.

Simulations of QEC codes under coherent errors can give a useful picture of the resilience against this kind of noise. Direct simulations of the general coherent noise model are limited by the exponential scaling of Hilbert space dimension with the number of qubits. This may be partially sidestepped using tensor network descriptions of the surface code, using which systems up to 153 qubits have been simulated~\cite{darmawanTensorNetworkSimulationsSurface2017}. The size of the system was not sufficient to establish a threshold, but it provided evidence that using the so-called Pauli twirl to approximate coherent errors as incoherent noise on the level of physical qubits underestimates the logical error rate.

A key advance for understanding the effect of coherent errors was the recent development of an algorithm capable of simulating a subset of coherent errors with effort that scales polynomially with the system size~\cite{bravyiCorrectingCoherentErrors2018}. The algorithm exploits a  representation~\cite{kitaevAnyonsExactlySolved2006, wenQuantumOrdersExact2003a} of the surface code in terms of Majorana fermions, and links this to coherent errors via the classically efficiently simulable \cite{terhalClassicalSimulationNoninteractingfermion2002} fermion linear optics (FLO) framework. 
By construction, the algorithm is limited to coherent errors acting as unitary rotations about one of the axes defined by the stabilizers, e.g., $U = \exp(i \eta Z)$, and it was developed for surface codes defined on a square lattice.

Here we describe a general approach  for representing surface codes with Majorana fermions on arbitrary planar graphs, including planar lattices, and show how the FLO-based algorithm can be adapted to this case.
For incoherent errors, it was found~\cite{fujiiErrorLossTolerances2012, rothlisbergerIncoherentDynamicsToric2012} that by changing the lattice geometry, one can trade off resilience against phase flips for resilience against bit flips. 
By applying our method to various lattices, and relating $Z$-rotations in one lattice to $X$-rotations in its dual lattice, we show that a similar trade-off is present for coherent errors as well, but now for $Z$- and $X$-rotations instead of $Z$- and $X$- (that is phase- and bit-) flips.

Furthermore, we study the distribution of states resulting from the application and correction of a coherent error and investigate whether, and if so in what sense, the logical-level noise decoheres, i.e., is approximable by a distribution of Pauli errors. We show that the answer depends on a graph classification that we establish. En route to our analysis of the  logical-level coherence, we also describe a coherent decoder that takes advantage of the deterministic nature of coherent errors. 

\section{Surface Code on General Lattices}
\label{sec:scongl}

Stabilizer codes, and as such surface codes, are constructed by defining a set of independent, mutually commuting, products $g_j$ of Pauli operators, in particular $g_j^2=I$ and $g_j\neq -I$~\cite{gottesmanStabilizerCodesQuantum1997, nielsenQuantumComputationQuantum2000}. The logical subspace is the subspace of the Hilbert space ``stabilized" by the $g_j$: $\ket{\psi}$ is in the codespace if $g_j\ket{\psi} = \ket{\psi}$ for all $j$. The condition $g_j\neq -I$ is required for the logical subspace to be nontrivial. In order to perform error correction with such a code, first each of the stabilizers is measured. The tuple $s$ of outcomes that is obtained is referred to as syndrome. From this syndrome, the decoder of the code computes a Pauli correction operation $C_s$ that brings the code back to a state in which all stabilizers measure $+1$, i.e., the logical subspace.

\begin{figure}
	\includegraphics{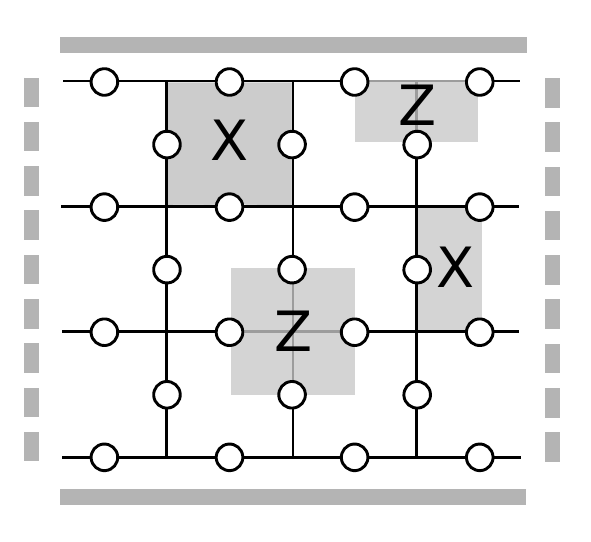}
	\caption{A small surface code on a square lattice. The white circles mark qubits. At each vertex of the lattice a $Z$-stabilizer is placed, acting on all adjacent qubits. On each plaquette, that is an area surrounded (or at the boundary partially surrounded) by links, an $X$- stabilizer is placed, acting on all qubits that are on its boundary. Examples of $X$- and $Z$-stabilizers are indicated as grey boxes. The code patch has two rough (dashed bars) and two smooth boundaries (solid bars). For both types of boundaries an appropriately truncated stabilizer is shown.	
	}
	\label{scbare}
\end{figure}

The surface code is a particular stabilizer code derived from the toric code~\cite{kitaevQuantumComputationsAlgorithms1997}. It is usually defined on a patch of a square lattice with a qubit placed on each of the links as shown in Fig.~\ref{scbare}. Each of the vertices are associated with a $Z$-stabilizer, $\prod_j Z_j$, where the product is taken over the qubits adjacent to the vertex. Conversely, each plaquette, that is a square surrounded by links, carries an $X$-stabilizer, $\prod_j X_j$, where the product is over the qubits on the plaquette boundary. In the toric code this pattern is placed on a torus or some other manifold without a boundary~\cite{kitaevQuantumComputationsAlgorithms1997}. The surface code, in contrast, is a planar construction based on a patch with boundaries~\cite{bravyiQuantumCodesLattice1998, fowlerSurfaceCodesPractical2012}.
In order to obtain a finite sized surface code, the pattern must be terminated.
The choice of boundaries determines the number of encoded logical qubits. The most often used boundaries are so-called rough boundaries and smooth boundaries. 
The rough boundaries are made of qubits on which only one (instead of two) $Z$-stabilizer acts. In terms of the lattice, they are stubs pointing out of the boundary of the patch, hence the name rough boundary. While the $Z$-stabilizers are unchanged compared to their bulk form at such a boundary, the $X$-stabilizers need to be modified due to the truncation of the plaquettes. 
Smooth boundaries are boundaries without such stubs. 
Now the $X$-stabilizers are unchanged compared to their bulk form, but the $Z$-stabilizers need to be modified. 
The boundary stabilizers are shown in Fig.~\ref{scbare}.
A patch with two rough and two smooth boundaries, in alternation, as shown in Fig.~\ref{scbare}, encodes one logical qubit. The logical $Z$-operator can be formed by a product of $Z$-operators acting on each of the qubits on one of the rough boundaries. Similarly, the logical $X$-operator can be constructed by a product of $X$-operators acting on each qubit on one of the smooth boundaries.

\begin{figure}
	\includegraphics[width=8.6cm]{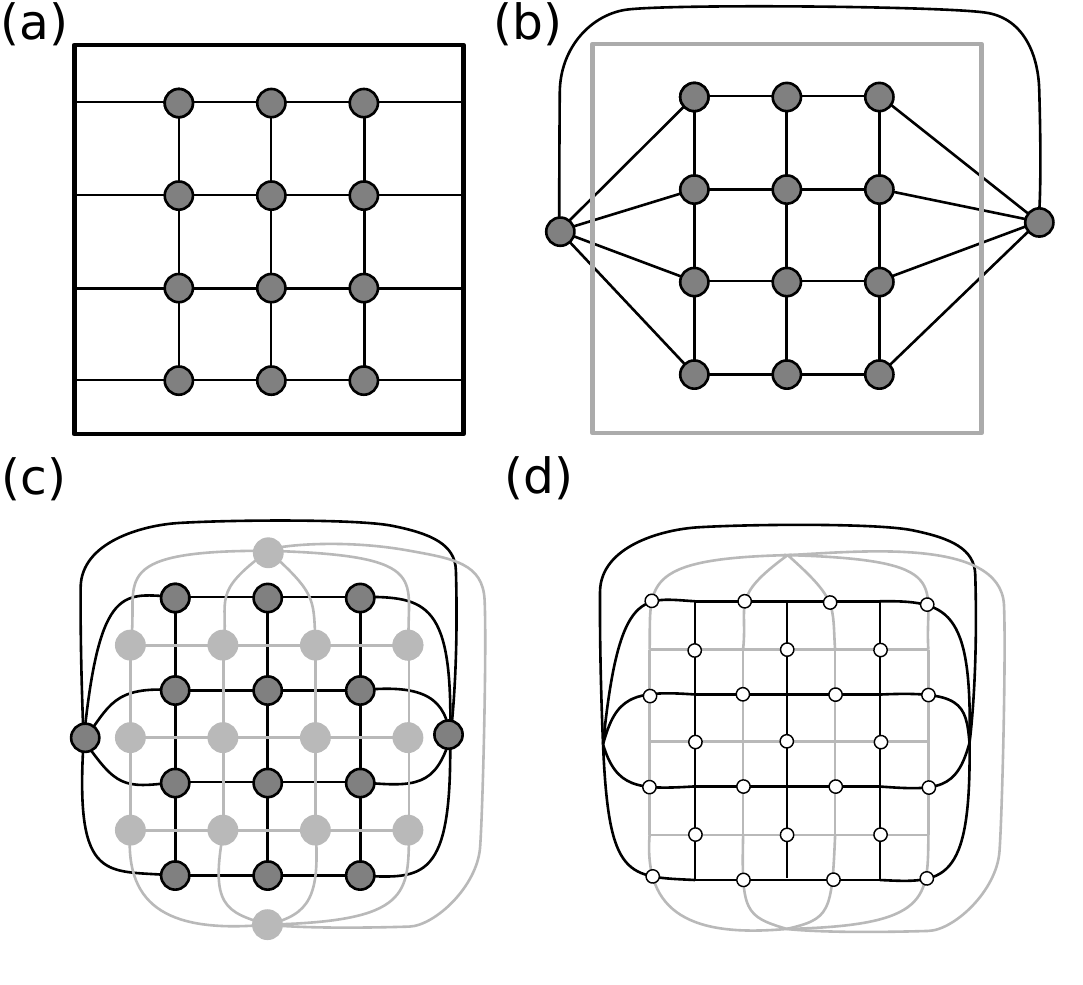}
	\caption{(a) The embedded planar graph of the $Z$-stabilizers for the setup equivalent to that in Fig.~\ref{scbare}. Grey circles mark the nodes of the graph that represent $Z$-stabilizers. (b) The same patch with added virtual stabilizers (see text). (c) The dual of the graph, forming the graph of $X$-stabilizers. Light grey circles mark the nodes of the dual graph that represent $X$-stabilizers. (d) The final code, made up of both the $X$- and $Z$-stabilizer graph and the qubits. 
	The qubits are shown as white circles. (To avoid clutter, the stabilizer nodes are not shown.)}
	\label{surfacesquare}
\end{figure}

To facilitate defining and describing surface codes on arbitrary planar graphs, we first  reformulate the above discussion in terms of graphs. We illustrate our considerations in Fig.~\ref{surfacesquare}.  For this construction, we start with a graph representing the $Z$-stabilizers, together with external edges that mark the rough boundaries [Fig.~\ref{surfacesquare}(a)].
To distinguish the two rough boundaries from each other and to keep track of them we add two connected virtual nodes, each connecting to all edges that belong to one of the rough boundaries as shown in Fig.~\ref{surfacesquare}(b). Using the virtual nodes, we can bring the external edges, which form the rough boundaries, to the conventional notion of graphs. This enables us to formulate the next steps in terms of standard graph operations. The virtual nodes themselves will not be translated into stabilizers of the code; instead, they can be used to define the logical operators, as we shall later explain.

From that graph we can obtain the graph of $X$-stabilizers by building the dual. The dual of a graph embedded in a surface is constructed by placing a node inside all faces that are formed by the edges of that graph and connecting two nodes if the faces they were placed on share a common edge. This is illustrated in Fig.~\ref{surfacesquare}(c). The graph that is obtained is the graph of $X$-stabilizers. It also contains two virtual nodes, defined as the nodes that are connected by the edge that is crossing the edge connecting the virtual $Z$-stabilizers. They are also not translated into stabilizers. Finally, the qubits are placed on the intersections of edges from the $X$- and $Z$-stabilizer graphs, where the intersection of the edges connecting virtual nodes is left out, resulting in a code patch shown in Fig.~\ref{surfacesquare}(d). Each stabilizer acts on all qubits it is directly connected to. The stabilizers thus defined are guaranteed to commute because a face always shares two edges with each of the vertices on its corner, therefore the resulting node in the $X$-stabilizer graph will share two qubits with the node from the $Z$-stabilizer graph. Since the overlap is on an even number of qubits, the corresponding operators commute.

Logical operators, i.e., Pauli products that commute with all stabilizers, but are independent of them,  can be obtained from the virtual nodes. Constructing an operator $X_\text{L}$ by building a product of $X$-operators over all qubits that are connected to one of the virtual nodes in the $X$-stabilizer graph produces an operator that commutes with all stabilizers for the same reasons the stabilizers commute with all other stabilizer, i.e., it overlaps on an even number of qubits with any of the $Z$-stabilizers
and trivially commutes with all $X$-stabilizers. The same holds for the operator $X'_\text{L}$ that can be obtained by choosing the other virtual node from the $X$-stabilizer graph, as well as for $Z_\text{L}$ and $Z'_\text{L}$ the operators obtained from placing $Z$-operators on the qubits attached to the virtual nodes of the $Z$-stabilizer graph.

This gives a total of four operators (one for each virtual node), however only two of them are independent: $X'_\text{L}$ can be obtained from $X_\text{L}$ by multiplication with all $X$-stabilizers and $Z'_\text{L}$ from $Z_\text{L}$ by multiplication with all $Z$-stabilizers.
Hence, we need to study only $X_\text{L}$ and $Z_\text{L}$. 
While $X_\text{L}$ and $Z_\text{L}$ commute with all stabilizers, they anticommute with each other: we did not include the qubit in the intersection of the edges between the virtual nodes, hence $Z_\text{L}$ and $X_\text{L}$ overlap only on a single qubit. Therefore, they form a pair of logical $X$- and $Z$-operators. We choose $Z_\text{L}$ to be the logical $Z$-operator and $X_\text{L}$ the logical $X$-operator.

\begin{figure}
	\includegraphics[width=8.6cm]{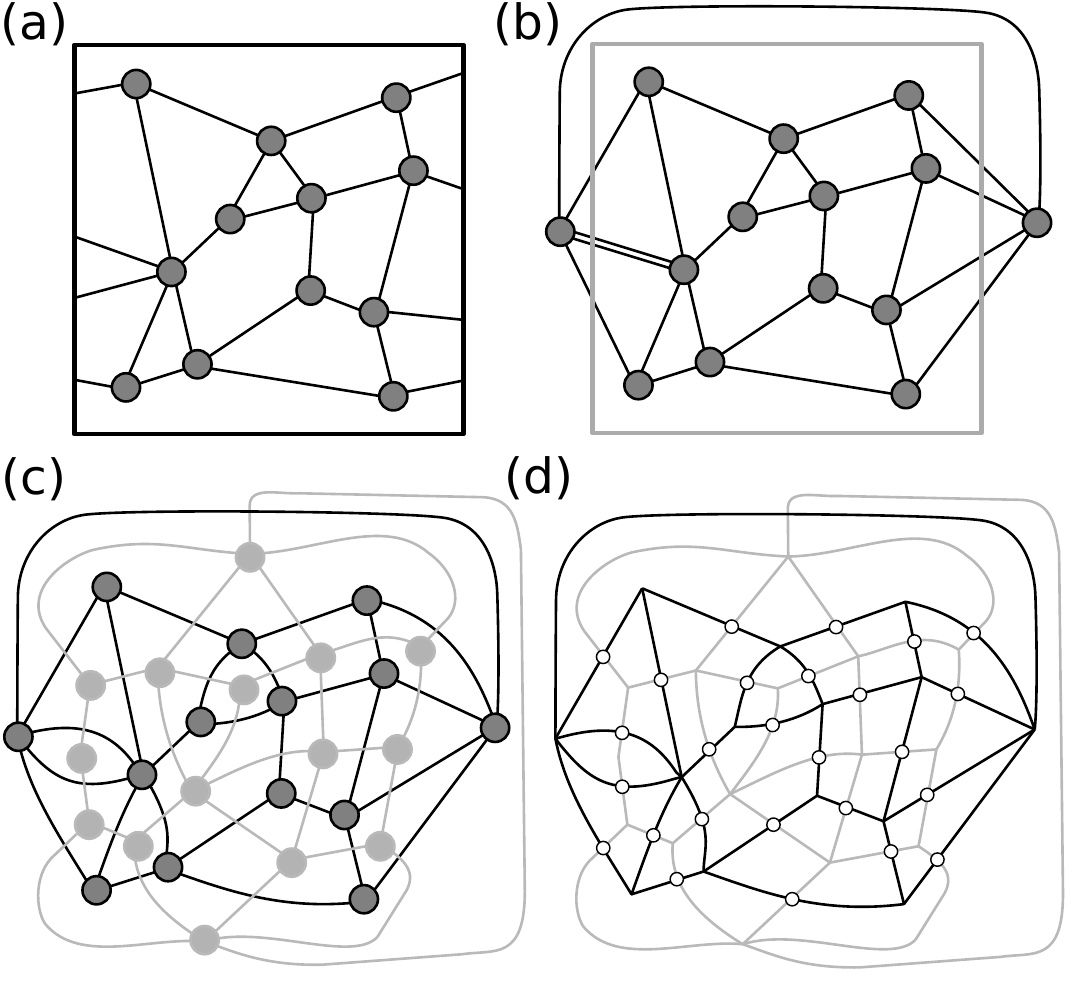}
	\caption{(a) An example of a general planar graph 	with two rough boundaries (left and right), defining the $Z$-stabilizers (dark grey circles) of the surface code patch. (b) The same graph with added virtual stabilizers to keep track of the code boundaries (see text). (c) In light grey, the dual graph, defining the $X$-stabilizers (light grey circles). (d) The qubits (white circles) placed on the resulting surface code patch.}
	\label{surfacecode}
\end{figure}

Applied to the square lattice, this construction recovers our previous discussion, as can be seen by comparing Figs.~\ref{scbare} and~\ref{surfacesquare}. However, it provides a framework for describing arbitrary planar graphs; an example is shown in Fig.~\ref{surfacecode} with the four panels describing the steps analogous to those in Fig.~\ref{surfacesquare}.

A surface code defined on a general planar graph shares many properties with the standard square-lattice surface code. Both are Calderbank-Shor-Steane \cite{steaneErrorCorrectingCodes1996} codes, which means that all stabilizers are either formed only by $X$- or only by $Z$-operations, a property we shall exploit to analyse the action of the error. Also, they can both be decoded by a minimum weight perfect matching (MWPM) algorithm \cite{dennisTopologicalQuantumMemory2002,fowlerPracticalClassicalProcessing2012, edmondsPathsTreesFlowers1965}.
Nevertheless, there are subtle differences, e.g., in the average stabilizer weight or the connectivity of the graph.  In particular, the connectivity has been shown to influence whether the code, under incoherent errors, is more resilient against bit or phase flips~\cite{fujiiErrorLossTolerances2012, rothlisbergerIncoherentDynamicsToric2012}.

\section{Error Models}

\newcommand{\numq}{N} 

In these planar graph surface codes, we shall study error channels 
\begin{equation}\label{eq:totalerror}
	\mathcal{E} = \bigotimes_{j=1}^\numq \mathcal{E}_j,
\end{equation}
where $\numq$
 is the number of qubits and $\mathcal{E}_j$ is a single-qubit error acting on qubit $j$. Our primary focus is the study of coherent errors of the form
\begin{equation}\label{eq:single_qubit_coh}
	\mathcal{E}_j[\rho] = \exp\left(i Z_j \eta_j \right) \rho \exp\left(-i Z_j \eta_j \right),
\end{equation}
where $\eta_j$ is a real parameter. 
The consideration of mere $Z$-rotations is linked to the FLO simulability of the system~\cite{bravyiCorrectingCoherentErrors2018}. We note, however, that due to the duality relation between the $X$- and $Z$-stabilizer graphs, we can also study $X$-rotations by exchanging the graphs for their duals.

To compare the effects of coherent errors to the incoherent case, we shall also study  Eq.~\eqref{eq:totalerror} with Eq.~\eqref{eq:single_qubit_coh} replaced by its Pauli twirl
\begin{equation}\label{eq:single_qubit_incoh}
	\mathcal{E}_j[\rho] = \rho \cos^2 \eta_j + Z_j \rho Z_j \sin^2 \eta_j.
\end{equation}
We shall be interested in studying what refinements of the finding of Ref.~\cite{darmawanTensorNetworkSimulationsSurface2017} that the Pauli twirl underestimates the coherent error may arise in more general lattices, and to assess how potential   trade-offs between resilience against $X$- and $Z$-errors compare in the coherent and incoherent~\cite{fujiiErrorLossTolerances2012} cases. 

\section{Quantum Error-Correction and its Characterization}
\label{sec:QECC}

To study the logical errors that arise, we apply $\mathcal{E}$ to the code followed by error correction $\mathcal{R}=\sum_s \mathcal{R}_s$ based on the MWPM decoder. Here $\mathcal{R}_s$ is the quantum operation of measuring syndrome $s$ followed by the application of the corresponding Pauli correction. (In Sec.~\ref{sec:averagelogic} we provide a more detailed description of the procedure.)

Inspired by Ref.~\cite{bravyiCorrectingCoherentErrors2018}, we shall investigate the logical error rate $p_\text{L}$ and the properties of the distribution of states after error correction.
A key difference between the square-lattice case~ \cite{bravyiCorrectingCoherentErrors2018} and codes  on general planar graphs is related to whether the weight of all $Z$-stabilizers (of $Z_\text{L}$) is even (odd). 
(The weight of a Pauli operator is the number of qubits on which it acts non-trivially.)
In Ref.~\cite{bravyiCorrectingCoherentErrors2018}, an alignment of the square lattice is chosen where all $Z$-stabilizers have even weight and $Z_\text{L}$ has odd weight. (This does not hold in the conventional orientation shown in Fig.~\ref{scbare} due to the boundary stabilizers.)
As explained in Ref.~\cite{bravyiCorrectingCoherentErrors2018}, a key consequence of this is that $\mathcal{R}_s\circ \mathcal E $
acts as a unitary channel on the logical qubit; the state $\rho_s$ arising after $\mathcal{R}_s\circ \mathcal E $ is
\begin{equation}\label{eq:unitarychannel}
\rho_s=\frac{1}{P_s}\mathcal{R}_s\circ \mathcal E [\rho]= \exp(i\theta_s Z_\text{L}) \rho \exp(-i\theta_s Z_\text{L}),
\end{equation}
where neither $\theta_s$ nor the probability $P_s$ of syndrome $s$ depend on the initial logical state $\rho$. 
Hence, both $p_\text{L}$ and the properties of the final state $\rho_s$ can be studied via a statistical analysis of $\theta_s$. 

For more general graphs, apart from some special cases, this property is absent and
$P_s$ depends on the initial logical state $\ket{\psi_\text{L}}$ of the code; 
the error correction process thus reveals information about $\ket{\psi_\text{L}}$. 
In studying the logical error rate,
we eliminate this $\ket{\psi_\text{L}}$ dependence by defining $p_\text{L}$ as the diamond-norm distance between the actions on the logical subspace of the identity and the average logical channel~\cite{rahnExactPerformanceConcatenated2002,bravyiCorrectingCoherentErrors2018}
\begin{equation}
	\Lambda_\text{L} [\rho] = \sum_s P_s \rho_s = \sum_s \mathcal{R}_s\circ \mathcal{E}[\rho] = \mathcal{R} \circ \mathcal{E}[\rho].
\end{equation}

We are also interested in the properties of final states $\rho_s$. To mitigate the $\ket{\psi_\text{L}}$ dependence of $P_s$ in this case, we adopt a statistical approach based on averaging with respect to a uniform distribution of $\ket{\psi_\text{L}}$ across the Bloch sphere. With $\ket{\psi_\text{L}}$ thus chosen randomly, the syndrome probability 
\begin{equation}
P_s=\text{Tr}\left(\mathcal{R}_s\circ \mathcal{E}[\rho]\right)\equiv P(s|\rho)
\end{equation}
must be viewed as the probability of $s$ conditioned on the initial state being $\rho=\ket{\psi_\text{L}}\bra{\psi_\text{L}}$. 
We shall be interested in the Bloch-sphere-averaged distance between $\rho_s$ and $\rho$. For a suitable (semi)metric
$\delta^2(\rho_s,\rho)$ on the space of logical states, this is
\begin{equation}
\langle\delta^2(\rho_s,\rho)\rangle_\Omega=\int_\Omega d\rho P(\rho|s) \delta^2(\rho_s,\rho),
\end{equation}
where $\Omega$ is the Bloch sphere and the conditional probability $P(\rho|s)$ enters because we are after the Bloch-sphere average given that the syndrome outcome is $s$. The combined Bloch-sphere and syndrome average is 
\begin{equation}\label{eq:avgdelta2}
\sum_s P(s)\langle\delta^2(\rho_s,\rho)\rangle_\Omega=\sum_s\int_\Omega d\rho P(s,\rho)  \delta^2(\rho_s,\rho),
\end{equation}
where $P(s)=\int_\Omega d\rho P(s,\rho)$ is a marginal of the joint syndrome-Bloch-sphere distribution $P(s,\rho)$. 
For computational convenience, for $\delta^2$ we shall use the square of the trace-norm distance
\begin{equation}
	\delta (\rho_s,\rho) = \sqrt{1 - |\braket{\psi_{\text{L}} | \psi_s}|^2},
\end{equation} 
between $\rho$ and $\rho_s=\ket{\psi_s}\bra{\psi_s}$.
That is, we consider the average infidelity conditioned on measuring syndrome $s$. Eq.~\eqref{eq:avgdelta2} thus gives the average infidelity to the identity of the average logical channel. 
For a discussion of the relation between the average infidelity and the diamond-norm distance see in Refs.~\cite{sandersBoundingQuantumGate2015,bealeQuantumErrorCorrection2018,beigiSimplifiedInstantaneousNonlocal2011, wallmanRandomizedBenchmarkingConfidence2014,iversonCoherenceLogicalQuantum2020}

\section{Majorana Graph}

To study the model introduced above we represent the surface code on a planar graph in terms of a corresponding Majorana fermion graph.  Our approach is 
based on that of Refs.~\cite{kitaevAnyonsExactlySolved2006, wenQuantumOrdersExact2003a,bravyiCorrectingCoherentErrors2018}; it proceeds by representing physical qubits in terms of Majorana fermions and a local constraint. 
In this way, the eigenstates of the surface code are described in terms of free-fermion eigenstates of a  quadratic commuting-dimer Majorana Hamiltonian projected to the physical, qubit, Hilbert space.

To obtain this representation each qubit $j$ is encoded in four Majorana fermions $c_{j1}$, $c_{j2}$, $c_{j3}$, $c_{j4}$. This is referred to as $C4$-encoding. The Majorana fermions $c_{jk}$ satisfy
\begin{equation}
c_{jk}^\dagger = c_{jk}, ~~~ \{c_{ik}, c_{jl}\} = 2 \delta_{ij} \delta_{kl}
\end{equation}
where $\{...\}$ denotes the anticommutator and $\delta_{ij}$ is the Kronecker delta. 
A conventional fermion, satisfying $\{d_{m}, d_{n}^\dagger\} = \delta_{mn}$, is built out of a pair of Majorana fermions via $d_m = c_{m1} + i c_{m2}$.
The four Majorana fermions for qubit $j$ thus correspond to two conventional fermions, hence a four-dimensional Hilbert space. 
To arrive at a two-dimensional Hilbert space encoding a qubit, we introduce the stabilizer $S_j = -c_{j1}c_{j2}c_{j3}c_{j4}$ and work in the subspace satisfying $S_j = 1$ for all qubits $j$. We shall refer to the $S_j$ as qubit stabilizers. 

In the $C4$-encoding, the Pauli operators on a qubit are given by Majorana bilinears, $X_j = i c_{j1} c_{j2}, Y_j = i c_{j1} c_{j3}, Z_j = i c_{j2} c_{j3}$. We shall call these bilinears Pauli dimers. They satisfy the commutation relation for Pauli operators and commute with the qubit stabilizer $S_j$. Since $C4$-encoded states are stabilized by $S_j$, there is for each Pauli dimer an equivalent Pauli dimer: $X_j = i c_{j2} c_{j3} S_j$, $Y_j = i c_{j2} c_{j4} S_j$, $Z_j = i c_{j4} c_{j1} S_j$.

\begin{figure}
	\includegraphics[width=8.6cm]{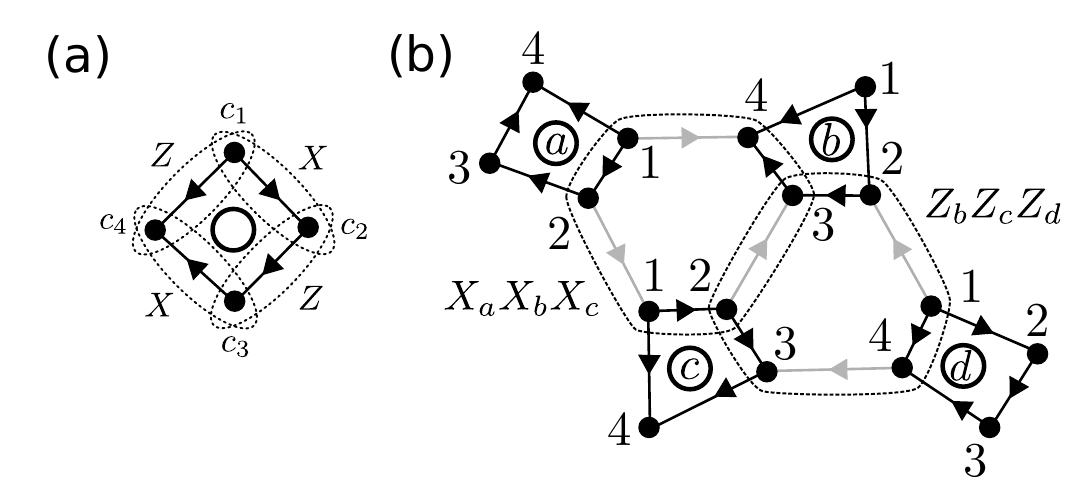}
	\caption{
		(a) Representation of a qubit in $C4$-encoding
		with four Majorana fermions $c_1, c_2, c_3, c_4$. The Pauli operators $X$ and $Z$ can be formed each by two equivalent dimers (pairs) of Majorana fermions. The order of the operators is indicated by arrows, e.g., $X = i c_1 c_2$ is represented by an arrow from $c_1$ to $c_2$.
		(b) Majorana graph for four qubits. A stabilizer with support on different qubits can be represented in terms of the Pauli dimers (black) of the qubits, but also using link dimers (grey) between the qubits. Link dimer orientations must be chosen such that to ensure equivalence to Pauli dimerization of stabilizers; this results in a Majorana graph where edges (Pauli and link dimers) have Kasteleyn orientation. 
	}
	\label{cfourencoding}
\end{figure}

It is beneficial to represent $C4$-encoded qubits in terms of a Majorana graph, as shown in Fig.~\ref{cfourencoding}. In this graph, nodes represent Majorana fermions and the edges between them represent bilinears $i c_{ik} c_{jl}$. The edges have an orientation indicated by arrows reflecting the operator order: for a bilinear $i c_{ik} c_{jl}$, the arrow points from fermion $c_{ik}$ to fermion $c_{jl}$. The Pauli dimers we shall use for a single qubit $j$ are those for $X_j$ and $Z_j$. The graph for a single qubit is shown in Fig.~\ref{cfourencoding} (a).

The $Z$- and $X$-stabilizers of the surface code involve products of Pauli operators from different qubits. Such products translate into products of Majorana operators which we have the freedom to reorder, provided we keep track of the signs. As illustrated in Fig.~\ref{cfourencoding} (b), this freedom can be used to change from Pauli dimers to ``link dimers", i.e., on the links between the qubits. 
In the example of Fig.~\ref{cfourencoding} (b), the stabilizer $X_aX_bX_c$ is rearranged
\begin{equation}
\begin{aligned}
X_a X_b X_c &= (i c_{a1} c_{a2}) (i c_{b3} c_{b4}) (i c_{c1} c_{c2}) \\
&= (i c_{a1} c_{b4}) (i c_{c2} c_{b3}) (i c_{a2} c_{c1}).
\end{aligned}
\end{equation}
Note that the rearrangement is performed in such a way that the dimer $i c_{c2} c_{b3}$ is also a part of the rearrangement of the stabilizer $Z_bZ_cZ_d$.

More generally, consider a stabilizer $g$ involving a product of $n$ Pauli operators, for which we can pick a dimer representations such that the Pauli dimers can be connected by additional edges such that the joint set of added edges and Pauli dimers form the boundary of a face in the Majorana graph. Then we can represent $g$ in terms of a clockwise product of Majorana fermions along the boundary of that face. Considering this operator order, each of the Pauli dimers in $g$ has the form $i s_k c_{\alpha_k} c_{\beta_k}$, where $\alpha_k$ and $\beta_k$ enumerate the double (i.e., qubit and Majorana) indices along the boundary of the face and $s_k=-1$ if this  operator order is opposite to that of the original Pauli dimer ($s_k=1$ otherwise). That is, 
\begin{equation}
	g = ( i s_1 c_{\alpha_1} c_{\beta_1}) (i s_2 c_{\alpha_2} c_{\beta_2}) \, ... \, (i s_n c_{\alpha_n} c_{\beta_n}).
\end{equation}
We now rearrange the product such that it is over link dimers. A simple rebracketing is sufficient for this for all but the first and the $n$-th Pauli dimer, 
\begin{equation}
	\begin{split}
	g = c_{\alpha_1} ( i s_1 c_{\beta_1} c_{\alpha_2} ) (i s_2 c_{\beta_2} c_{\alpha_3}) \, ... \\ ... \, ( i s_{n-1} c_{\beta_{n-1}} c_{\alpha_n}) (i s_n c_{\beta_n}).
	\end{split}
\end{equation}
To form the link dimer $i c_{\beta_n} c_{\alpha_1}$, however, $c_{\alpha_1}$ has to be commuted through an odd number of Majorana fermions yielding
\begin{equation}\label{eq:Kasteleyn}
	\begin{split}
	g = -( i s_1 c_{\beta_1} c_{\alpha_2} ) (i s_2 c_{\beta_2} c_{\alpha_3}) \, ... \\ ... \, ( i s_{n-1} c_{\beta_{n-1}} c_{\alpha_n}) (i s_n c_{\beta_n} c_{\alpha_1}).
	\end{split}
\end{equation}
Provided we absorb this additional sign in the orientation of one of the link dimers, we now find that the stabilizer is expressed as a product over these. Since all but one of these have a corresponding Pauli dimer with the same orientation, the total number of clockwise oriented edges (i.e., Pauli and link dimers) around the face of the Majorana graph is odd. By the same logic, the same holds for each of the faces that represent a stabilizer in the graph. The Majorana graph thus has an orientation in which all faces have an odd number of clockwise oriented edges: a so-called Kasteleyn orientation.

The free fermion state underlying the description of the surface code emerges from the observation that the entire set of stabilizer generators can be rearranged in the way described above and thereby be described in terms of mutually commuting link dimers. 
(These dimers, however, do not commute with the qubit stabilizers $S_j$, highlighting the fact that the surface code eigenstates are not free-fermion states, but projections thereof.)
For the square lattice, this is shown in Refs. \cite{kitaevAnyonsExactlySolved2006, wenQuantumOrdersExact2003a, bravyiCorrectingCoherentErrors2018}. In the following we shall describe an algorithm with which one can construct Majorana graphs for surface codes on arbitrary planar graphs.
We shall illustrate our algorithm using the $Z$-stabilizer graph in Fig.~\ref{surfacecode}(a) and the corresponding qubit graph in Fig.~\ref{surfacecode}(d).

\begin{figure}
	\includegraphics[width=8.6cm]{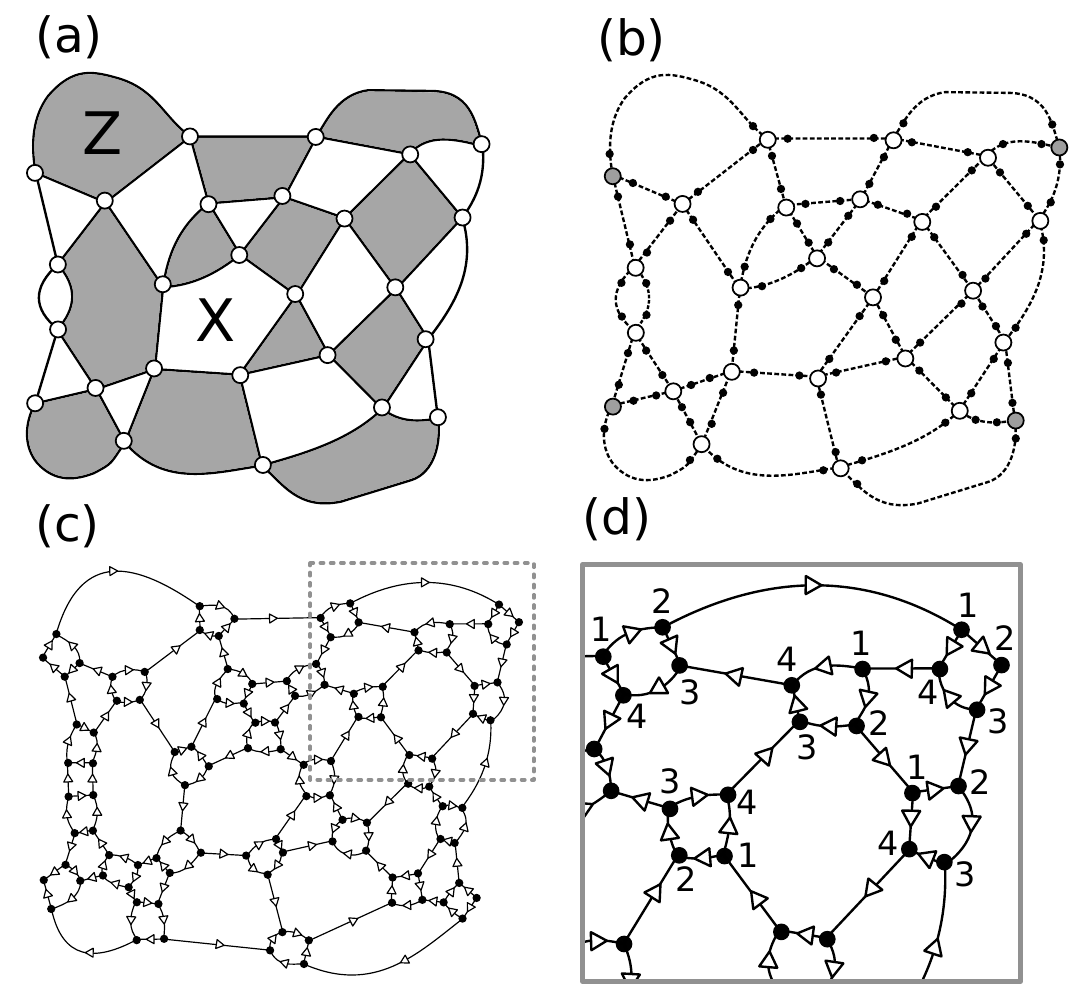}
	\caption{(a) The construction of graph $\mathcal{G}$ with one face per stabilizer for the surface code in Fig.~\ref{surfacecode}. (b) The edges of $\mathcal{G}$ can be used as Majorana dimers for the $C4$-encoding of the surface code. This results in four Majorana fermions per qubit except for the four qubits in the corners (grey). (c) A Kasteleyn orientation of the dimers. (d) Zoom from panel (c) showing the $C4$-encoding of the individual qubits (cf. Fig~\ref{cfourencoding}).}
	\label{fig:majoranagraph}
\end{figure}

We start the construction of the Majorana graph $\mathcal{M}$ with the qubit graph $\mathcal{Q}$, the graph containing both qubits and stabilizers as vertices and connecting each stabilizer to the qubits they are acting on. The construction of $\mathcal{Q}$ starting from any initial planar graph of $Z$-stabilizers is described in Sec.~\ref{sec:scongl} [an example is shown in Fig.~\ref{surfacecode}(d)].
To construct $\mathcal{M}$, we will construct one intermediate graph $\mathcal{G}$ by taking all qubits from $\mathcal{Q}$ and connecting them if this can be done without crossing any edge in $\mathcal{Q}$. Fig.~\ref{fig:majoranagraph}(a) shows the graph $\mathcal{G}$ obtained in this way from  $\mathcal{Q}$ in Fig.~\ref{surfacecode}(d). 
The graph $\mathcal{G}$ has the property that it contains a face for every stabilizer generator and the operator it represents acts on the qubits on the boundary of that face. 
Additionally, it has the property that every qubit, apart from those at the corners,   has four edges connected to it. Qubits at the corners are different because these miss the link to the qubit that was not inserted because it would have been at the intersection of edges connecting virtual stabilizers.

To obtain the Majorana graph $\mathcal{M}$ we place two Majorana fermions on each of the edges in $\mathcal{G}$ and associate each of the fermions to one of the qubits at the edge's ends. Using this method, we associate with each qubit the same number of Majorana fermions as the qubit has edges connected to it. Since in the graph $\mathcal{G}$ each of the vertices, except the four on the corners of the graph, is connected with four edges, all qubits, except the qubits in the corner, have four Majorana fermions associated to them. For the example system we are considering, this stage of the construction is shown in Fig.~\ref{fig:majoranagraph}(b). In order to be able to encode all qubits in the $C4$-encoding we add one additional Majorana fermion to each of the qubits at the corners of the code, such that these additional fermions are in none of the faces of the graph $\mathcal{G}$. This gives the complete set of vertices for the graph $\mathcal{M}$.

To construct the edges of $\mathcal{M}$, we first add an edge between two of its vertices if they were on the same edge in $\mathcal{G}$;  these edges will form the link dimers. We also add the edges associated to the $C4$-encoding of the qubits by adding the edges forming a face around each qubit; these edges will form the Pauli dimers. 
This completes the construction of the (thus far unoriented) edges of $\mathcal{M}$.

Next, we have to associate some more structure to $\mathcal{M}$:
We must assign (i) a Kasteleyn orientation [cf. Fig.~\ref{cfourencoding} and under Eq.~\eqref{eq:Kasteleyn}] and (ii) a placement of Pauli dimers (i.e., a numbering $c_{j1},\ldots,c_{j4}$ of Majorana fermions) around each qubit such that the clockwise Majorana product around each face of $\mathcal{G}$ encodes the correct stabilizer. 
That such structure exists can be seen as follows. First, we consider the Pauli dimers. In the bulk of the code, the stabilizers surrounding a qubit alternate between $X$ and $Z$, since qubits
are placed on the intersection of an edge of the $X$-stabilizer graph and an edge of the $Z$-stabilizer graph. Similarly, the dimer representation of the Pauli operators in the $C4$-encoding alternates between $X$- and $Z$-dimers [cf. Fig.~\ref{cfourencoding} (a)]. 
Therefore, we can always arrange Pauli dimers such that each of them is adjacent to the stabilizer face to which it contributes.
(On the edge of the code, there are less than four surrounding stabilizers,  however, the existing adjacent stabilizers already specify the placement of $X$- and $Z$-dimers).
In a convention where $X$-dimers are oriented clockwise [Fig.~\ref{cfourencoding} (a)], 
there are two possible orientations for each qubit: we can choose which of the $Z$ Pauli dimers is oriented anti-clockwise. We can pick any of the two. This defines the Pauli dimer part of the Majorana graph, including the orientation of the faces around each of the qubits. 

To complete the structuring of $\mathcal{M}$, we must orient the link dimers such that globally a Kasteleyn orientation is obtained. To this end, we can use that each of the so far unoriented faces (edges) in $\mathcal{M}$ is associated to a face (edge) in the graph $\mathcal{G}$. Then, for any face in $\mathcal{G}$, we count the number $n$ of clockwise-oriented Pauli dimers surrounding the corresponding face in $\mathcal{M}$. If $n$ is odd, we have to orient the edges of this face of $\mathcal{G}$ such that an even number of edges are clockwise (and vice versa for $n$ even). 
In this way, it is sufficient to 
find an orientation of $\mathcal{G}$ such that each face has the parity of clockwise oriented edges as determined by $n$ before. To produce this orientation for $\mathcal{G}$, we can proceed similarly to the first steps of the FKT algorithm \cite{kasteleynGraphTheoryTheoretical1967}. By orienting $\mathcal{M}$'s link dimers  according to the orientation obtained for $\mathcal{G}$, we have obtained a Kasteleyn orientation of $\mathcal{M}$. The resulting graph and orientations for our example are shown in Fig.~\ref{fig:majoranagraph}(c, d).

\begin{figure}
	\includegraphics[width=8.6cm]{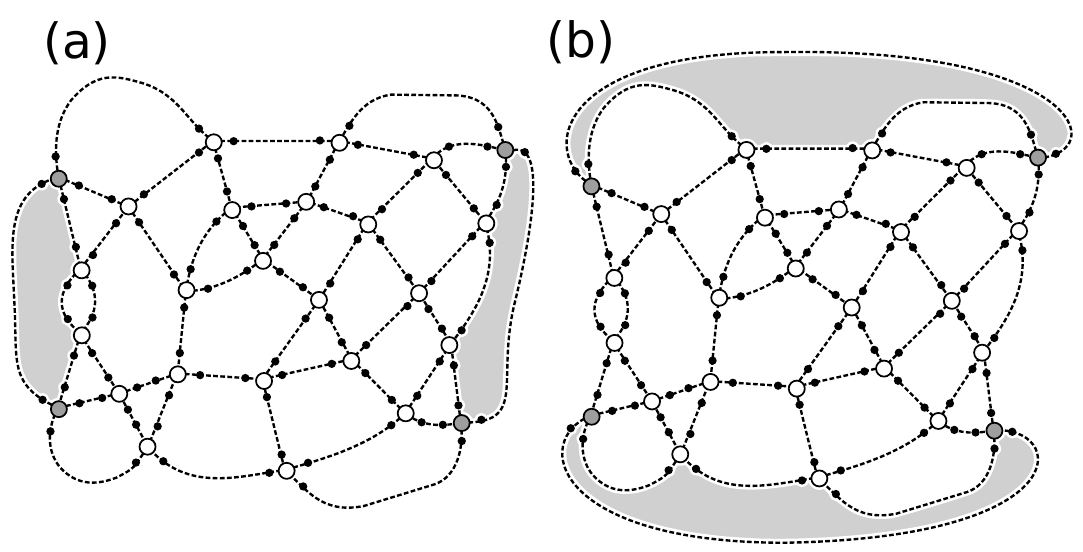}
	\caption{Majorana encoding of logical operators. A logical operator of the code translates to an additional face in $\mathcal{M}$ bounded by a link dimer formed by two of the four initially unpaired corner Majorana fermions. Subfigure  (a) shows $\mathcal{M}$ with the two link dimers that form both realizations of $Z_\text{L}$ (each dimer corresponds to one of the two virtual $Z$-stabilizers of Sec.~\ref{sec:scongl}). The additional faces of $\mathcal{M}$ for including these dimers are shown in grey.
	Subfigure (b): link dimers for $X_\text{L}$, showing again the two additional faces.}
	\label{fig:logicoperator}
\end{figure}

The logical state of the surface code patch is defined by the state of the qubit encoded in the four unpaired corner Majorana fermions~\cite{bravyiCorrectingCoherentErrors2018}. This becomes clear
when we consider the Majorana encoding of the logical operators. Following steps analogous to Eq.~\eqref{eq:Kasteleyn}, the Majorana encoding of either of the logical operators requires a new link dimer connecting two of the four corner fermions (as shown in Fig.~\ref{fig:logicoperator}); the new link dimer has orientation such that the resulting new face (which corresponds to a virtual stabilizer for $\mathcal{G}$) has an odd number of edges pointing clockwise. In the initial state in which all stabilizers measure $+1$ all equivalent realizations of a logical operator must have the same expectation value. Therefore, we have to add such faces for both realizations (corresponding to both of the virtual stabilizers) of the logical operator. 
By stabilizing the state that is encoded in the Majorana graph with a logical operator, we fix the logical state of the code to be in a $+1$ eigenvalue of that logical operator. We thereby fix the code to be in either the $\ket{0_\text{L}}$ state by choosing to stabilize with $Z_\text{L}$ or in the $\ket{+_\text{L}}$ state by using $X_\text{L}$. To initialize the code in the $\ket{Y_\text{L}}$ state, we can pair up fermions from diagonally opposite ends of the code patch.

\section{FLO Simulation}
\label{sec:FLO}

In the following we describe how to use the methods introduced in Ref.~\cite{bravyiCorrectingCoherentErrors2018} to sample from the distribution of syndromes and how to compute, given a syndrome $s$, the overlaps 
\begin{equation}
	\bra{\pm_\text{L}} C_s \exp i  \pmb{\eta} \pmb{Z} \ket{+_\text{L}}, \bra{\pm_\text{L}} C_s \exp i \pmb{\eta} \pmb{Z} \ket{Y_\text{L}},
\end{equation}
where $C_s$ is the Pauli correction for syndrome $s$,
$\pmb{\eta} = (\eta_1, \eta_2, \ldots, \eta_\numq), \pmb{Z} = (Z_1, Z_2, \ldots, Z_\numq)$, and $\pmb{\eta} \pmb{Z}$ is the scalar product between the two.
From these overlaps, the quantities characterizing the error correction process [cf. Sec.~\ref{sec:QECC}] can be extracted using Monte Carlo simulation, as shown in Secs.~\ref{sec:averagelogic} and \ref{sec:residualcoherence}. 

To perform these operations, we use the framework of fermion linear optics (FLO). Within this framework we have access to the following operations:
\begin{itemize}
\item Initializing a dimer in the $+1$ eigenstate.
\item Applying the unitary operation $R = \exp(\eta c_i c_j)$ with an arbitrary real $\eta$.
\item Projectively measuring a dimer operator, with or without post selection.
\end{itemize}
These are operations that maintain the property of a state to be a fermionic Gaussian state, which can be exploited to simulate their actions efficiently~\cite{terhalClassicalSimulationNoninteractingfermion2002}.

The limitation of the FLO algorithm is that it cannot treat quartic products of Majorana operators such as those in the qubit stabilizers $S_j$. To bypass this problem, each qubit is projectively measured in the $\ket{\pm}$-basis; we shall see that this allows working with objects involving Majorana bilinears. 
Although this makes it impossible to evaluate the $Z$-stabilizers, such evaluation is not needed: since we apply only $Z$-rotations we know that none of the $Z$-stabilizers could have been flipped.

To sample from the syndrome distribution, we sample from the eigenvalues $m_a$ of single-qubit Pauli operators $X_a$; the eigenvalues of $X$-stabilizers can be computed from $m_a$ classically. The probability for measuring $m_a$ requires performing three steps on each qubit $a$: first switch to fermions and project into the $C4$-encoding using $(1 + S_a) / 2$, then apply the coherent error $U_a = \exp{I \eta_a Z_a}$, and finally apply the projector $(1 + m_a X_a) / 2$. Since $Z_a$ commutes with the qubit stabilizer $S_a$ we can perform the rotation first; a further reordering of the Majorana fermions gives
\begin{equation}
	P_a(m_a) = \frac{1}{2} (1 + m_a X_a) \frac{1}{2} (1 + m_a S_a X_a) \exp(I \eta_a Z_a),
\end{equation}
in terms of which the joint probability of measuring \mbox{$\pmb{m} = (m_1, m_2, \ldots m_n)$} is the expectation value of $\prod_{a=1}^\numq P_a(m_a)$ with respect to a Gaussian state.
Note that computing this joint distribution requires only rotations and measurements with post selection, both involving dimers only, allowing a computation using the FLO operations introduced above. 
However, this by itself does not offer a route to efficiently sample from the exponentially many outcomes $\pmb{m}$. 
Ref.~\cite{bravyiCorrectingCoherentErrors2018} showed how one may do this qubit by qubit, thus breaking the sampling down to a repeated sampling from just two states. 
The success of this approach hinges on choosing a correct order in which to measure the qubits: The order must be such that the graph $\mathcal{G}$ stays connected when removing, after every measurement, the qubit that was measured. Such an ordering can be obtained for our graphs by performing a breadth-first search through the graph;  the obtained order can be used in reverse.

From the sampled $\pmb{m}$, we compute the syndrome $s$ classically using suitable products of $m_a$ for the corresponding $X$-stabilizers. From $s$, the decoder produces the correction $C_s$. Since, by construction, no $Z$-stabilizer is flipped, the correction contains only operators that correct $X$-stabilizers: $C_s$ contains only $Z$-operators. Using this, we  define $\pmb{\eta}_s$ according to 
\begin{equation}
	\exp{i \pmb{\eta}_s \pmb{Z}} = C_s \exp{i \pmb{\eta Z}}, 
\end{equation}
that is, we absorb the correction operations into the parameters of the coherent rotations.

The quantity we aim to compute 
is $\bra{+_\text{L}} \exp i \pmb{\eta}_s \pmb{Z} \ket{+_\text{L}}$. We can expand $\ket{+_\text{L}}$ in the computational basis; it is given by the sum over the set $\mathcal{L}$ of all computational basis states that satisfy all $Z$-stabilizers,
\begin{equation}
	\ket{+_\text{L}} = | \mathcal{L} |^{-1/2} \sum_{x \in \mathcal{L}} \ket{x}.
\end{equation}
Furthermore, 
\begin{multline}
	\bra{+_\text{L}} \exp i \pmb{\eta}_s \pmb{Z}  \ket{+_\text{L}} = | \mathcal{L} |^{-1} \sum_{x \in \mathcal{L}} \sum_{y \in \mathcal{L}}  \bra{y} \exp i \pmb{\eta}_s \pmb{Z} \ket{x}\\
	=| \mathcal{L} |^{-1}\sum_{y \in \{0,1\}^\numq}\sum_{x \in \mathcal{L}}   \bra{y} \exp i \pmb{\eta}_s \pmb{Z} \ket{x}\\
	=2^{\numq/2}| \mathcal{L} |^{-1} \sum_{x \in \mathcal{L}}\bra{+^{\otimes \numq}} \exp i \pmb{\eta}_s \pmb{Z} \ket{x}\\
	=2^{\numq/2}| \mathcal{L} |^{-1/2}\bra{+^{\otimes \numq}} \exp i \pmb{\eta}_s\ket{+_\text{L}},
\end{multline}
where in the second line we used that $\exp i \pmb{\eta}_s \pmb{Z}$ is diagonal in the computational basis to replace $\sum_{y \in \mathcal{L}}$ by a summation over all computational basis states, and in the third line we used that 
$2^{\numq/2}\ket{+^{\otimes \numq}}=\sum_{y \in \{0,1\}^\numq}\ket{y}$. 
Hence, 
\begin{equation}\label{eq:overlap_prob}
	|\bra{+_\text{L}} \exp i \pmb{\eta}_s \pmb{Z}  \ket{+_\text{L}}|^2 = M^{-2}| \bra{+^{\otimes \numq}} \exp i \pmb{\eta}_s \pmb{Z}  \ket{+_\text{L}}|^2.
\end{equation}
with $M=2^{-\numq/2}| \mathcal{L} |^{1/2}$. Eq.~\eqref{eq:overlap_prob} is a constant $M^{-2}$ times the probability to measure the outcome $\ket{+}$ for all qubits. Hence it can be computed using the FLO algorithm, this time without sampling, to find the probability for the outcome $\pmb{m} = (1, 1, ...)$. To eliminate the factor $M$ we can build the ratio
\newcommand{\symqs}{q_{s}}
\newcommand{\symrs}{r_{s}}
\begin{equation}
\begin{split}
	\symqs = \frac{|\bra{-_\text{L}} C_s \exp i \pmb{\eta}_s \pmb{Z} \ket{+_\text{L}}|^2}{|\bra{+_\text{L}} C_s \exp i \pmb{\eta}_s \pmb{Z} \ket{+_\text{L}}|^2} = \\ = \frac{|\bra{+^{\otimes \numq}} Z_\text{L} \exp i \pmb{\eta}_s \pmb{Z}  \ket{+_\text{L}}|^2}{|\bra{+^{\otimes \numq}} \exp i \pmb{\eta}_s \pmb{Z}  \ket{+_\text{L}}|^2}.
\end{split}
\end{equation}
For the simulation, the operator $Z_\text{L}$ can be absorbed into $\pmb{\eta}$ the same way we absorbed $C_s$.

In a similar fashion, we can compute the ratio
\begin{equation}
	\symrs = \frac{|\bra{-_\text{L}} C_s \exp i \pmb{\eta}_s \pmb{Z} \ket{Y_\text{L}}|^2}{|\bra{+_\text{L}} C_s \exp i \pmb{\eta}_s \pmb{Z} \ket{Y_\text{L}}|^2},
\end{equation}
i.e., the same expectation values but starting with the $\ket{Y_\text{L}}$ state. This can be done by initializing the simulation in a different state such that the logical state is given by $\ket{Y_\text{L}}$.

\section{Average Logical Channel}
\label{sec:averagelogic}

In the following, we explain how to obtain the full action of the average logical channel
from the observables $\symqs, \symrs$ that are accessible via the FLO simulation. 

To this end, we first need a description of the recovery procedure $\mathcal{R}$. The recovery scheme consists of two steps. First, all stabilizers are measured; this projects the state into one syndrome $s$. This projection is performed by the projector $\Pi_s$. Next, depending on the syndrome $s$, the decoder chooses a correction operation $C_s$. The combined recovery is given by
\begin{equation}
	\mathcal{R}[\rho] = \sum_s C_s \Pi_s \rho \Pi_s C_s^\dagger.
\end{equation} 
Since $C_s$ maps between the space in which the stabilizers have the syndrome $s$ and the logical subspace we can represent $\Pi_s = C_s \Pi_0 C_s^\dagger$, where $\Pi_0$ denotes the projection into the logical subspace. Using this relation and the fact that the corrections $C_s$ are Pauli operators and thus satisfy $C_s = C_s^\dagger$, the operation $\mathcal{R}$ can be expressed as
\begin{equation}
	\mathcal{R}[\rho] = \sum_s \Pi_0 C_s \rho C_s \Pi_0.
\end{equation}
The correction $\mathcal{R}$ together with the error $\mathcal{E}$ is
\begin{equation}
	\mathcal{R} \circ \mathcal{E}[\rho] = \sum_s \Pi_0 C_s \exp(i \pmb{\eta Z})  \rho \exp(-i \pmb{\eta Z}) C_s \Pi_0.
\end{equation}
This operation maps any state of the logical subspace back to the logical subspace. Within that subspace, it is the average logical channel
\begin{equation}\label{eq:LambdaL}
	\Lambda_\text{L} [\rho] = \mathcal{R} \circ \mathcal{E}[\Pi_0 \rho \Pi_0],
\end{equation} 
where we introduced $\Pi_0$ to remind that we view $\Lambda_\text{L} [\rho]$ as a quantum channel on the logical subspace.

Since we have an algorithm to sample from the distribution of syndromes, we study the action corresponding to an individual syndrome $s$:
\begin{equation}
	\mathcal{R}_s\circ \mathcal{E}[\Pi_0 \rho\Pi_0] = D_s \rho D_s^\dagger,
\end{equation}
where we introduced $D_s = \Pi_0 C_s \exp(i \pmb{\eta Z}) \Pi_0$. 
We consider the action of $D_s$ in the logical subspace. $D_s$ commutes with the logical $Z_\text{L}$ operator, therefore, $D_s$ is diagonal in the $Z_\text{L}$-basis and hence can be represented as 
\begin{equation}\label{eq:asbs}
	D_s =  \operatorname{Diag}(a_s, b_s) = a_s \pi_0 + b_s \pi_1,
\end{equation}
with $a_s, b_s \in \mathbb{C}$ and $\pi_i$ the projector on the logical states $\ket{i_\text{L}}\bra{i_\text{L}}$, $i\in {0,1}$. In terms of $a_s$ and $b_s$, we have
\begin{multline}
	\mathcal{R}_s\circ \mathcal{E}[\Pi_0 \rho\Pi_0] = |a_s|^2 \pi_0 \rho \pi_0 + |b_s|^2 \pi_1 \rho \pi_1 \\ + a_s \overline{b}_s \pi_1 \rho \pi_0  + \overline{a}_s b_s \pi_0 \rho \pi_1,
\end{multline}
where bar indicates complex conjugation.
The action of $\Lambda_\text{L}$ follows from $\sum_s\mathcal{R}_s$. We find
\begin{equation}
	\Lambda_\text{L} [\rho]  = \alpha \pi_0 \rho \pi_0 + \beta \pi_1 \rho \pi_1 + \gamma \pi_0 \rho \pi_1 + \overline{\gamma} \pi_1 \rho \pi_0 \label{eq:firstgammaexpansion},
\end{equation}
with
\begin{equation}
	\alpha = \sum_s |a_s|^2, ~~~ \beta = \sum_s |b_s|^2, ~~~ \gamma = \sum_s \gamma_s, ~~~ \gamma_s = a_s \overline{b}_s.
\end{equation}
Since $\Lambda_\text{L}$ is trace preserving, we have $\alpha = 1$ and $\beta = 1$. Thus, the entire action of $\Lambda_\text{L}$ is encoded in the single complex parameter $\gamma$. Furthermore, 
\begin{equation}\label{eq:LmI}
(\Lambda_\text{L}-\mathbbm{1})[\rho]=(\gamma-1) \pi_0 \rho \pi_1 + (\overline{\gamma}-1) \pi_1 \rho \pi_0,
\end{equation}
from which, by the proportionality of Eq.~\eqref{eq:LmI} to the action of a unitary channel minus the identity, we read off~\cite{johnstonComputingStabilizedNorms2009,sandersBoundingQuantumGate2015,bravyiCorrectingCoherentErrors2018}
the diamond-norm \cite{kitaevQuantumComputationsAlgorithms1997} distance 
\begin{equation}
	p_\text{L}=\norm{\Lambda_\text{L} - \mathbbm{1}}_\diamond = |\gamma - 1|.
\end{equation}

We wish to estimate $\gamma$ using Monte Carlo simulation. Using the FLO approach, we are able to sample from the distribution of syndromes starting from the initial state $\ket{+_\text{L}}$. The syndrome probability is
\begin{equation}\label{eq:Ps+}
	P_s^+ = \norm{ D_s \ket{+_\text{L}} }^2.
\end{equation}
To estimate $\gamma$ we seek a quantity $c_s$ accessible from the simulation such that 
\begin{equation}
P^+_s c_s = \gamma_s. \label{eq:montecarloform}
\end{equation} 
In this way, the Monte Carlo average $\sum_s P^+_s c_s=\gamma$. 

Using the simulation algorithm introduced in Sec.~\ref{sec:FLO}, we have access to
\begin{equation}
	\symqs = \frac {|\bra{+_\text{L}} Z_\text{L} D_s \ket{+_\text{L}}|^2} {|\bra{+_\text{L}} D_s \ket{+_\text{L}}|^2} ~~ \text{and} ~~ \symrs = \frac{|\bra{+_\text{L}} Z_\text{L} D_s \ket{Y_\text{L}}|^2}{|\bra{+_\text{L}} D_s \ket{Y_\text{L}}|^2}.
\end{equation}
By expanding both expressions in $a_s$ and $b_s$ and reordering we find the relations
\begin{equation}
	\symqs = \frac{P^+_s - \Re \gamma_s}{P^+_s + \Re \gamma_s}, ~~ \symrs = \frac{P^+_s - \Im \gamma_s}{P^+_s + \Im \gamma_s},
\end{equation}
which imply
\begin{equation}
	\Re \gamma_s = \frac{1 - \symqs}{1 + \symqs} P^+_s, ~~~ \Im \gamma_s = \frac{1 - \symrs}{1 + \symrs} P^+_s.
\end{equation}
Conveniently, both expressions match the form of Eq. (\ref{eq:montecarloform}). 
Therefore, we can approximate $\gamma$ using a Monte Carlo approximation of the sum
\begin{equation}
	\gamma = \sum_s P^+_s c_s = \sum_s P^+_s \left( \frac{1 - \symqs}{1 + \symqs} + \frac{1 - \symrs}{1 + \symrs} i\right).
\end{equation}

\section{Threshold}
\label{sec:threshold}

\begin{figure*}
	\includegraphics[width=\textwidth]{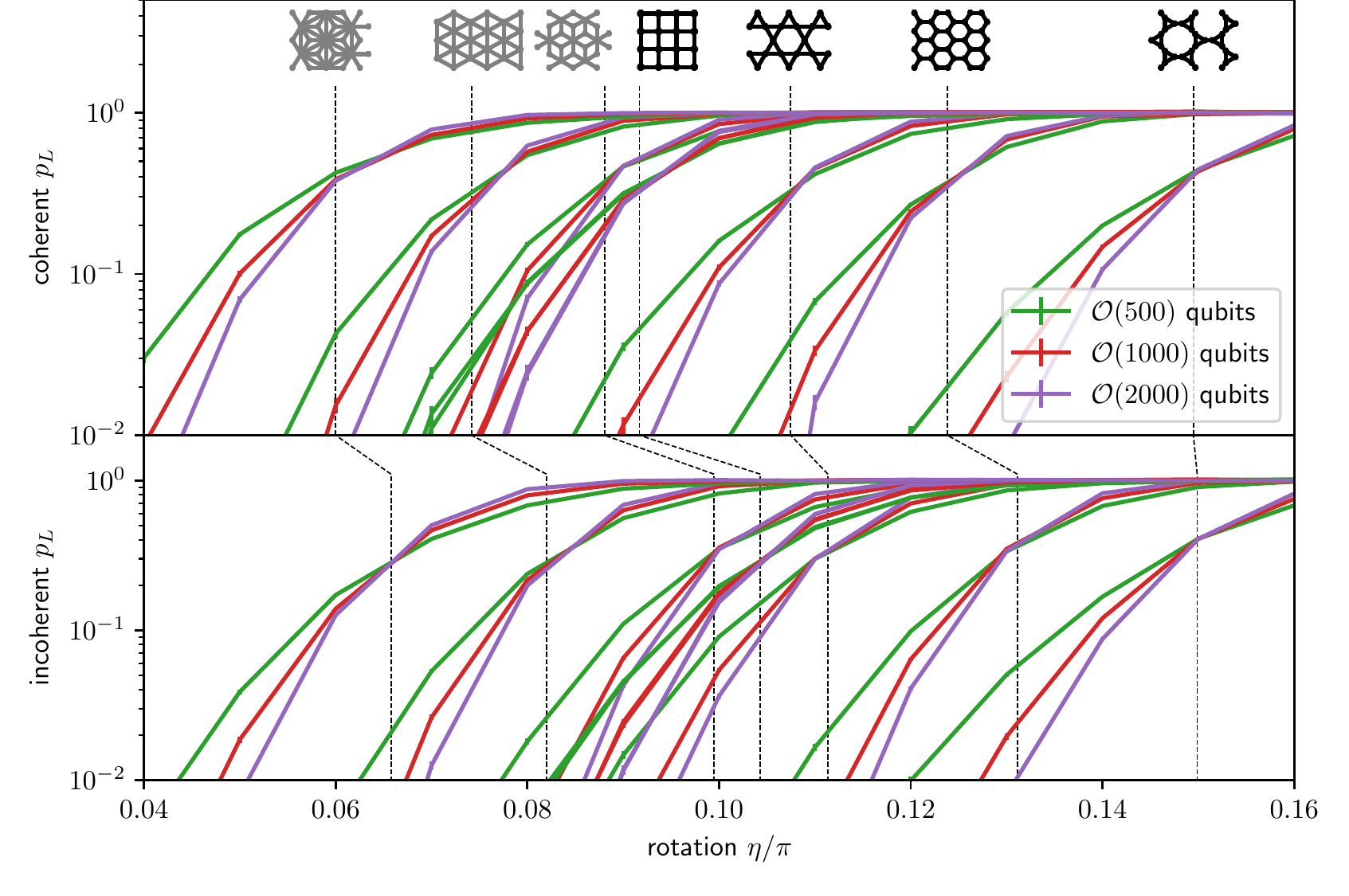}
	\vspace{-1cm}
	\caption{The logical error rate $p_\text{L}$ for the coherent and Pauli-twirled error models versus the angle $\eta$ for three different code sizes of each lattice. (The argument of $\mathcal{O}$ indicates the approximate number $N$ of qubits; the concrete value of $N$ depends on the lattice.)
	The dashed lines indicate the thresholds obtained by fitting a finite-size scaling ansatz~\cite{wangSurfaceCodeQuantum2011}. Above the threshold, we sketch the graphs of the $X$-stabilizers. They are, from left to right, dual of tri-hex, dual of hexagonal, dual of kagome, square, kagome, hexagonal, and tri-hex.}
	\label{fig:threshold}
\end{figure*}

Simulations of surface codes on various lattices have shown that the thresholds of the codes depend significantly on the connectivity of the lattice~\cite{fujiiErrorLossTolerances2012, rothlisbergerIncoherentDynamicsToric2012}. In the following, we study lattices with different connectivity under coherent and incoherent errors.

In the choice of lattices, we follow Ref.~\cite{fujiiErrorLossTolerances2012} and perform simulations for the square, kagome, hexagonal, $(3,12^2)$ (triangle-hexagonal, also  referred to as tri-hex) lattices, and their duals. For the square lattice, we study codes with distances 25 (625 qubits), 37 (1369 qubits), 49 (2401 qubits) and for the other lattices we study system sizes with a comparable number of qubits.

\begin{figure}
	\begin{center}
	\vspace{-0.2cm}
	\includegraphics[width=8.6cm]{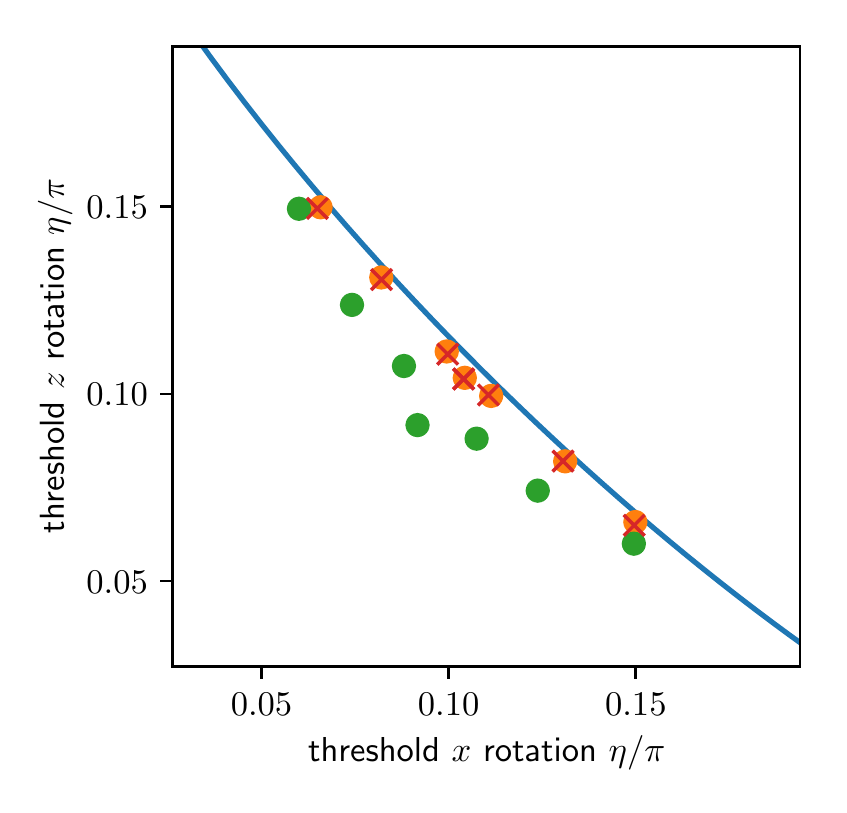}
	\end{center}
	\vspace{-1cm}
	\caption{The thresholds
	for $X$-rotations versus the thresholds for $Z$-rotations, for both the coherent (green) and Pauli-twirled (orange) error models, for the lattices indicated in Fig.~\ref{fig:threshold}.  (The thresholds obtained in Ref.~\cite{fujiiErrorLossTolerances2012} are shown in red.) The blue line indicates the bound Eq.~\eqref{eq:bound}.}
	\label{fig:thresholddependency}
\end{figure}

For each surface code we perform two simulations: one for coherent errors [i.e., with $\mathcal{E}$ using Eqs.\eqref{eq:totalerror} and \eqref{eq:single_qubit_coh}] and one with incoherent errors using the Pauli twirl [i.e., with $\mathcal{E}$ using Eqs.\eqref{eq:totalerror} and \eqref{eq:single_qubit_incoh}].
For simplicity, we apply the same error to all of the qubits, $\eta_j=\eta$. 
For both error models and all lattices we first simulate an initial overview spanning from $\eta = 0.4 \pi$ to $\eta = 1.6 \pi$ in $0.1 \pi$ steps with 10000 Monte Carlo samples for the coherent error and 40000 Monte Carlo samples for the incoherent error. The results are shown in Fig.~\ref{fig:threshold}. We then estimate the thresholds $\eta_\text{th}$ by first estimating their position from this overview, and then performing a simulation in $0.01 \pi$ steps around the estimated position and fit a finite-size scaling ansatz~\cite{wangSurfaceCodeQuantum2011}. 
Our threshold estimates are shown in Fig.~\ref{fig:threshold} and Fig.~\ref{fig:thresholddependency}.
We find that for these lattices the coherent thresholds are consistently higher than the incoherent ones (or are at best comparable to them as for the tri-hex lattice). 
Our results also indicate that for $\eta\lesssim \eta_\text{th}$ the logical error rate decreases with code distance slower for coherent than for incoherent errors. 

The trade-off between resilience against bit- and phase- flips that is obtainable in the incoherent (twirled) case is reflected by the thresholds approaching~\cite{fujiiErrorLossTolerances2012} the bound
\begin{equation}\label{eq:bound}
	R \leq 1 - h(p_x) - h(p_z)
\end{equation}
for zero asymptotic encoding rate $R\rightarrow 0$. (For our case of a single encoded qubit, $R=1/\numq$.) Here $h$ is the binary Shannon entropy and  $p_x$ and $p_z$ are the probabilities of $X$- and $Z$-flips on  individual physical qubits~\cite{gottesmanStabilizerCodesQuantum1997}. 
In our case, the thresholds $p_{z,\text{th}}$ are parametrized by $\eta$, i.e., $p_{z,\text{th}} = \sin^2 \eta_\text{th}$ and $p_{x,\text{th}}$ is obtained by lattice duality. In Fig.~\ref{fig:thresholddependency} we visualize this bound.

The results show that the trade-off between resilience against bit- and phase- flips translates, for coherent errors, to a trade-off between resilience against $X$- and $Z$- rotations. In Ref.~\cite{fujiiErrorLossTolerances2012} it is argued that the trade-off for incoherent errors is present because it is easier for the MWPM decoder to match up syndromes in a sparse graph. It is reasonable to assume that a similar effect is also causing the trade-off in the coherent case. However, considering Fig.~\ref{fig:thresholddependency}, unlike for the incoherent thresholds, there does not appear to be a universal curve delineating this trade-off for coherent thresholds.

\begin{figure}
	\begin{center}
	\includegraphics[width=8.6cm]{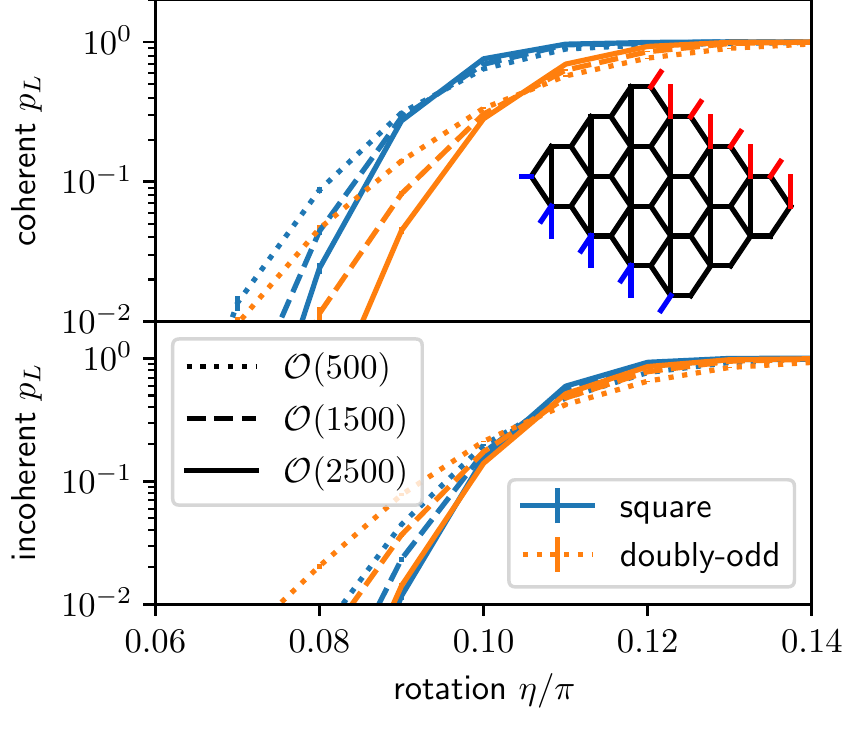}
	\end{center}
	\vspace{-1cm}
	\caption{A comparison between the logical error rate $p_\text{L}$ for the doubly-odd and square lattice. (A small patch of the doubly-odd lattice shown as inset.) The upper graph shows $p_\text{L}$ for coherent errors, the lower shows $p_\text{L}$ for incoherent errors. The simulations for the square lattice are the same used for Fig. \ref{fig:threshold}. The doubly-odd lattice is simulated for system sizes of 437, 1365, and 2805 qubits.}
	\label{fig:dodperformance}
\end{figure}

To provide further evidence for the absence of such universal curve,
we construct a lattice that is self-dual and therefore can be directly compared to the square lattice, i.e., its thresholds for $X$- and $Z$-errors are equal by design. We call this lattice the ``doubly-odd" lattice; it has faces with 3 and 5 vertices (see Fig.~\ref{fig:dodperformance} inset). 
We compare the results for this lattice, both for coherent and incoherent errors, to the square-lattice case in Fig.~\ref{fig:dodperformance}.
The results show that the coherent threshold for the doubly-odd lattice is significantly higher than that for the square lattice. For incoherent errors, however, the thresholds for the two lattices are very close, consistently with the observation of Ref.~\cite{fujiiErrorLossTolerances2012} that most surface codes with a MWPM decoder perform very close to the bound Eq.~\eqref{eq:bound}.

\section{Final-State Distribution}
\label{sec:residualcoherence}

To get further insights into the properties of the states after error correction, we study the action of the error and correction process conditioned on the individual syndromes. The final states have certain properties that are dependent on properties of the stabilizer group. It turns out that the parity of the stabilizers is of central importance. 
We shall generalize the property~\cite{bravyiCorrectingCoherentErrors2018} that for codes which have only even-weight $Z$-stabilizers together with an odd-weight logical $Z_\text{L}$-operator, a coherent $Z$-error followed by a correction acts as a unitary operation. 
We will assess the different symmetries that are present in the lattices we considered above and determine the consequences for the final-state distributions. Note that we are now investigating properties of the error correction process based on properties of the $Z$-stabilizer graph, while the argument that it is easier to correct errors in sparse graphs was based on the $X$-stabilizer graph.

We start by considering an individual syndrome $s$ that is corrected with the operator $C_s$ that is made up only of $Z$-operators. In the following, we denote a string of $Z$-operators by $Z(b)$, where $b$ is a bit-string of length $\numq$ whose value is one (zero) for qubits on which $Z(b)$ acts nontrivially (trivially). In particular, $Z(0)=I$. Expanding both the error and the correction using this representation yields
\begin{equation}
	C_s = Z(h_s), ~~ \exp(i \pmb{\eta Z}) = \sum_{g \in \mathcal{B}} c_g i^{\norm{g}} Z(g),
	\label{eq:err:expansion}
\end{equation} 
where $h_s$ is the bit-string encoding the correction operation, $c_g$ are real coefficients, $\norm{g}$ denotes the Hamming weight of the bit-string $g$, and $\mathcal{B}$ is the set of all length-$\numq$ bit-strings.  
Therefore~\cite{bravyiCorrectingCoherentErrors2018},
\begin{equation}
D_s =  \sum_{g \in \mathcal{B}} c_g i^{\norm{g}} \Pi_0 Z(h_s \oplus g) \Pi_0, \label{eq:err:expansion2}
\end{equation}
where $\oplus$ is addition modulo $2$. 

\begin{figure}
	\includegraphics[width=8.6cm]{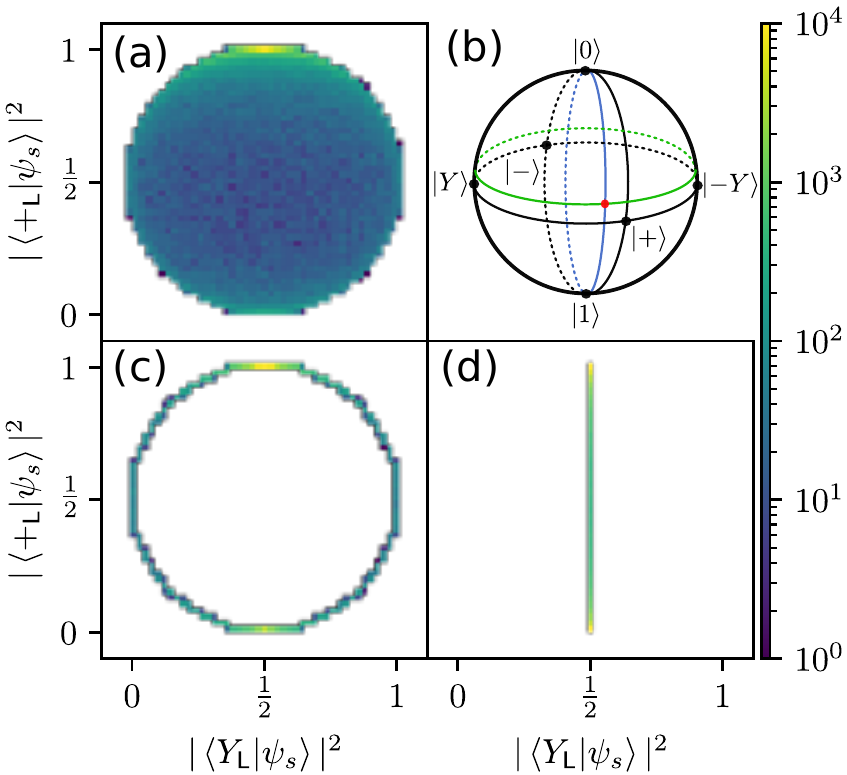}
	\vspace{-0.5cm}
	\caption{Panels (a), (c), (d):
	The projection of 100000 sampled states $\ket{\psi_s}$ resulting after error and correction operation with initial  state $\ket{+_\text{L}}$. Distribution for a surface code on the (a) hexagonal lattice; (c) square lattice; (d) kagome lattice.
	The size of the system is for all examples chosen such that the number of qubits is around 500.
	(b) The Bloch sphere and an example of an initial state  (red dot).
	The green (blue) circle marks the possible final states in lattices with even-weight $Z$-stabilizers and  odd-weight (even-weight) $Z_\text{L}$. 
	}
	\label{fig:diffin}
\end{figure}

As discussed in Sec.~\ref{sec:averagelogic}, $D_s$ is diagonal in the logical space of the code. Therefore, we can represent it as
\begin{equation}
	D_s = k_s \Pi_0 + l_s \Pi_0 Z_\text{L},
\end{equation}
with suitable coefficients $k_s$ and $l_s$. When expressing $k_s$ and $l_s$ as a sum over contributions from Eq.~(\ref{eq:err:expansion2}), $k_s$ is formed by terms in which $Z(h_s \oplus g)$ acts trivially on the logical space, i.e., when $Z(h_s \oplus g)$ is within the $Z$-stabilizer group. Conversely, $l_s$ is formed by  those terms in which $Z(h_s \oplus g)$ acts as the logical $Z_\text{L}$ operator, i.e., those corresponding to $Z_\text{L}$ times a stabilizer. Hence, 
\begin{equation}\label{eq:Bsum}
\begin{aligned}
	k_s &= \sum_{g \in \mathcal{B} ~\text{with}~ g \oplus h_s \in \mathcal{A}} c_{g} i^{\norm{g}},
	 \\
	 l_s &= \sum_{g \in \mathcal{B} ~\text{with}~ g \oplus h_s \in \mathcal{A} \oplus l} c_{g} i^{\norm{g}},
\end{aligned}
\end{equation}
where  $\mathcal{A}$ is the set of all bit-strings corresponding to operators in the $Z$-stabilizer group, $l$ is the bit-string that encodes $Z_\text{L}$, and  $\mathcal{A} \oplus l$ denotes the set $\{ a \oplus l | a \in \mathcal{A}\}$. Since $g$ runs over all bit-strings, we can convert Eq.~\eqref{eq:Bsum} into sums over $\mathcal{A}$, 
\begin{equation}\label{eq:shiftedsum}
\begin{aligned}
	k_s &= \sum_{a\in \mathcal{A}} c_{a \oplus h_s} i^{\norm{a \oplus h_s}},
	 \\
	 l_s &= \sum_{a \in \mathcal{A}} c_{a \oplus h_s \oplus l} i^{\norm{a \oplus h_s \oplus l}}.
\end{aligned}
\end{equation}

We next study the complex phase of $k_s$ and $l_s$. In general, $\mathcal{A}$ contains both even- and odd-weight bit-strings and we cannot make a definite statement. If, however, all bit-strings in $\mathcal{A}$ have even weight, the exponent of $i$ has the same parity for all the terms in each sum in Eq.~\eqref{eq:shiftedsum}. This constrains $k_s$ and $l_s$ to be either real or imaginary, with their relative phase set by $l$. Therefore, we can distinguish between two families: codes in which all $Z$-stabilizers are of even weight and codes that do not satisfy this condition.

In the case that not all $Z$-stabilizers are of even weight we can make no further statements
about the distribution. In fact, checking the distribution that is obtained for a code based on a $Z$-stabilizer graph defined on a hexagonal lattice, i.e., a system in which the majority of the $Z$-stabilizers are of weight 3,
we find [Fig.~\ref{fig:diffin}(a)] that at least with some probability all parts of the Bloch sphere of final states are reached.

For those lattices for which every $Z$-stabilizer has even weight, we can identify the relative complex phase between $l_s$ and $k_s$. For this phase, the parity of $h_s$ is irrelevant since, whether it is odd or even, it affects both $l_s$ and $k_s$ the same way. However, the weight of $Z_\text{L}$  affects only $l_s$. Hence, in the family in which all $Z$-stabilizers are of even weight we have two subfamilies, those with even-weight $Z_\text{L}$ and those with odd-weight $Z_\text{L}$.

\begin{figure*}
	\includegraphics[width=17.2cm]{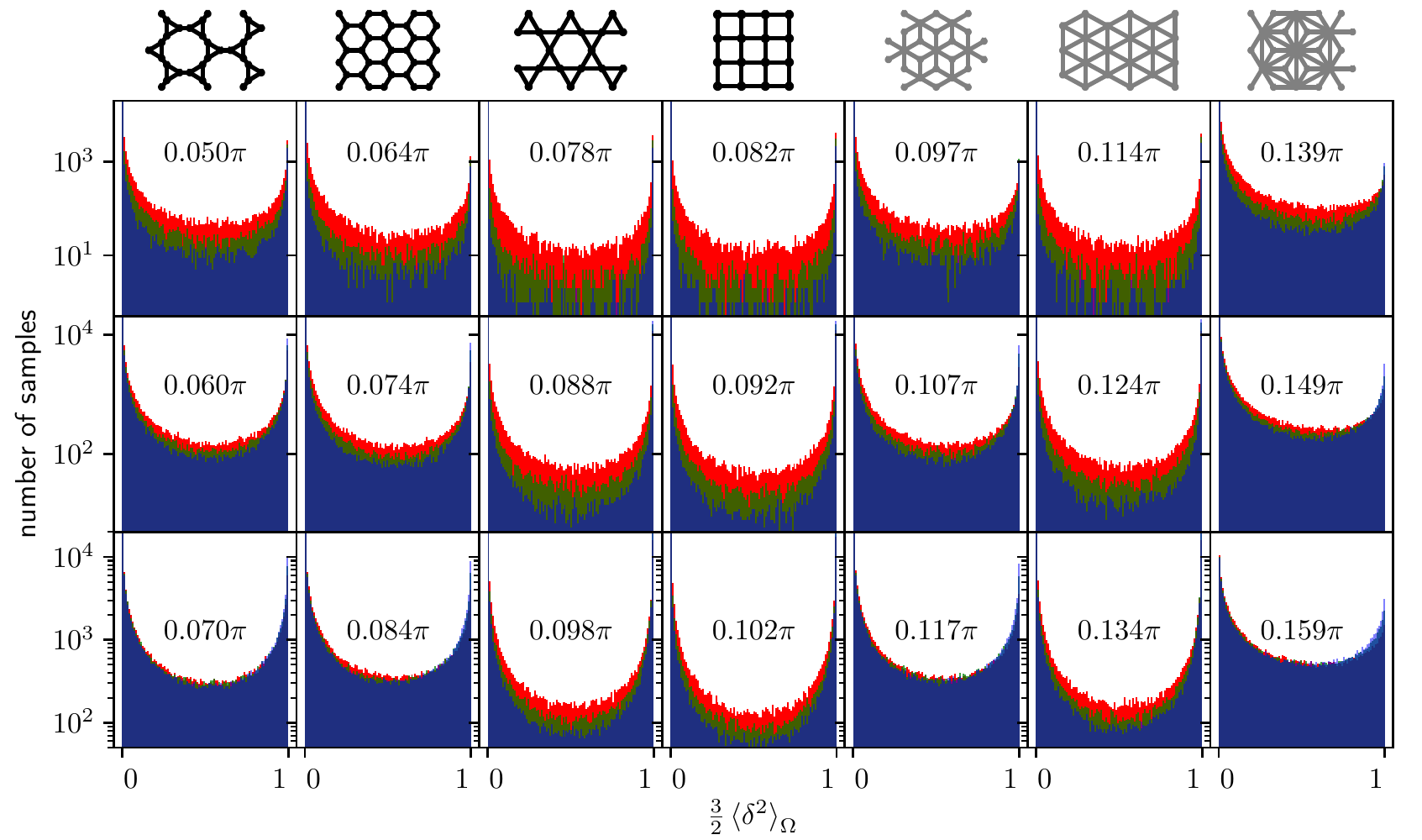}
	\caption{Distributions of 100000 samples of $\langle \delta^2 \rangle_\Omega$ for the nine lattices in Fig.~\ref{fig:threshold}. We give a sketch of the corresponding $Z$-stabilizer graph above the distributions. (These are the duals of the graphs in Fig.~\ref{fig:threshold}.) The values of the coherent noise parameter $\eta$ are shown above the distributions; they are chosen as $\eta\approx \eta_\text{th}$, i.e., approximately at the threshold (middle row), $\eta\approx \eta_\text{th}-0.01\pi$ (top row), and $\eta\approx \eta_\text{th}+0.01\pi$ (bottom row). Samples are taken for three different systems sizes $\mathcal{O}(500)$ qubits (red), $\mathcal{O}(1500)$ qubits (green), and $\mathcal{O}(2500)$ qubits (blue).}
	\label{fig:distr}
\end{figure*}

The case that $Z_\text{L}$ is of odd weight is already explored in Ref.~\cite{bravyiCorrectingCoherentErrors2018}: we have $\Im{k_s} = 0$ and $\Re{l_s} = 0$ and 
\begin{equation}
	D_s = \sqrt{P_s} U_s, ~~ P_s = |k_s|^2 + |l_s|^2, ~~ U_s = \frac{k_s}{\sqrt{P_s}} + \frac{l_s Z} { \sqrt{P_s}},
\end{equation}
where $U_s$ is a unitary operator. 
That is, the state $\rho_s$ satisfies Eq.~\eqref{eq:unitarychannel}. 
In consequence, the logical state is constrained to a circle on the Bloch sphere that is parallel to the $XY$-plane [Fig.~\ref{fig:diffin}(b)].
The distribution of states that can be obtained starting from the $\ket{+_\text{L}}$ state is shown in Fig.~\ref{fig:diffin}(c).

If $Z_\text{L}$ is of even weight, we have $\Im{k_s} = 0$ and $\Im{l_s} = 0$. The operation $D_s$ in that case is given by the, unusual, real combination of the identity and $Z_\text{L}$. If $|k_s| > |l_s|$,
\begin{equation}\label{eq:Ds1}
	D_s =  (|k_s| - |l_s|) + 2 |l_s| \left\{ \begin{array}{cc}   \pi_0 & \text{if} ~ l_s k_s > 0, \\ \pi_1 & \text{else,} \end{array} \right.
\end{equation}
and if $|l_s| > |k_s|$,
\begin{equation}\label{eq:Ds2}
	D_s =   Z_\text{L} (|l_s| - |k_s|) + 2  Z_\text{L} | k_s| \left\{ \begin{array}{cc} \pi_0 & \text{if} ~ l_s k_s > 0, \\  \pi_1 & \text{else.} \end{array} \right.
\end{equation}
Eqs.~\eqref{eq:Ds1} and \eqref{eq:Ds2} show that syndrome measurements 
reveal information about the $Z_\text{L}$-polarization of the initial logical state $\ket{\psi_\text{L}}$.
They also imply that the final state after the action of $D_s$ lies on the circle spanning $\ket{0_\text{L}}$, $\ket{1_\text{L}}$, and $\ket{\psi_{\text{L}}}$. This is also  illustrated in Fig.~\ref{fig:diffin}(b), and a numerical example is shown in Fig.~\ref{fig:diffin}(d).

This leaves us with three classes of lattices to build $Z$-stabilizer graphs, and we have examples for each:
\begin{itemize}
	\item containing odd-weight $Z$-stabilizers: tri-hex, dual of tri-hex, hexagonal and dual of kagome,
	\item all even-weight $Z$-stabilizers, odd-weight $Z_\text{L}$: square, dual of hexagonal, 
	\item all even-weight $Z$-stabilizers, even-weight $Z_\text{L}$: kagome.
\end{itemize}

To compare the final-state distributions for the different lattices, we proceed as explained in Sec.~\ref{sec:QECC}. 
We have $\ket{\psi_s}\!\!=D_s\!\ket{\psi_{\text{L}}}/\sqrt{P(s|\rho)}$ with $P(s|\rho)\!=\!\braket{\psi_{\text{L}}|D_s^\dagger D_s|\psi_{\text{L}}}$ (recall, $\rho=\ket{\psi_{\text{L}}}\bra{\psi_{\text{L}}}$).
Hence, 
\begin{equation}
	\delta^2 (\rho_s,\rho) = 1 -   \frac{|\braket{\psi_{\text{L}} |D_s| \psi_\text{L}}|^2}{\braket{\psi_{\text{L}} |D_s^\dagger D_s| \psi_\text{L}}} . \label{eq:expdeltasqr}
\end{equation}
To prepare for the Bloch-sphere average, we parameterize
\begin{equation}
	\ket{\psi_{\text{L}}} = \ket{\theta, \phi} = \cos(\theta / 2) \ket{0_\text{L}} + \sin (\theta / 2) e^{i \phi} \ket{1_\text{L}}.
\end{equation}
After some manipulation we find
\begin{equation}
\begin{aligned}
	\label{eq:avg}
	\delta^2 (\rho_s,\rho) & = \frac{|\braket{-_\text{L}|D_s|+_\text{L}}|^2 \sin^2 \theta}{P(s|\rho)},
\end{aligned}
\end{equation}
where
\begin{equation}\label{eq:prob}
	P(s|\rho) = \frac{1}{2}\left[|a_s|^2 + |b_s|^2 + (|a_s|^2 - |b_s|^2) \cos \theta\right]
\end{equation}
due to Eq.~\eqref{eq:asbs}. 
Using $P(\rho|s)=P(s|\rho)P(\rho)/P(s)$, the Bloch-sphere average is
\begin{equation}
\begin{aligned}\label{eq:Blochavg}
	\langle\delta_s^2\rangle_{\Omega} &\equiv \int_{\Omega}d\rho P(\rho|s)\delta^{2}(\rho_{s},\rho) \\&=   \frac{|\braket{-_\text{L}|D_s|+_\text{L}}|^2}{P(s)} \int_{\Omega}d\rho P(\rho)\sin^{2}\theta \\ & = \frac{2}{3} \frac{|\braket{-_\text{L}|D_s|+_\text{L}}|^2}{P(s)} = \frac{1}{3} \left( 1 - \frac{ \Re \gamma_s}{ P(s)} \right),
\end{aligned}
\end{equation}
where
we used that for $P(\rho)$ uniformly distributed over the Bloch sphere, $\int_{\Omega}d\rho P(\rho)\sin^{2}\theta=2/3$. Note that
\begin{equation}
P(s)=\int_{\Omega}d\rho P(s|\rho)P(\rho)=\frac{|a_{s}|^{2}+|b_{s}|^{2}}{2}=P_s^+,
\end{equation}
the probability in Eq.~\eqref{eq:Ps+}. Hence, Eq.~\eqref{eq:Blochavg} is entirely in terms of quantities that can be extracted from the FLO-based simulation.

We sample from $\langle\delta_s^2\rangle_{\Omega}$
for three different values of the error parameter $\eta$ for each lattice: approximately the threshold value ($\eta\approx\eta_\text{th}$) and $\eta\approx\eta_\text{th}\pm 0.01\pi$. The results are shown in Fig.~\ref{fig:distr}. For all simulations, we observe that 
$\langle\delta_s^2\rangle_{\Omega}$
has sharp peaks around $2/3$ and $0$; 
these values correspond to $\ket{\psi_s}=Z_\text{L}\!\ket{\psi_\text{L}}$ and $\ket{\psi_s}=\ket{\psi_\text{L}}$, respectively.
This shows that the coherent $Z$-rotations for the codes at the distances we study can be well approximated by a distribution of Pauli errors. However, in contrast to the effect of an incoherent error, each state $\rho_s$ after the operation $\mathcal{R}_s\circ \mathcal{E}$ is still a pure state; we get a mixed state only if the information of the syndrome outcome $s$ is deleted (i.e., only for the output of $\Lambda_\text{L}$).

The distributions show patterns characteristic of the lattice geometry. In particular, while below the threshold all lattices have a distribution that is increasingly concentrated on $0$ and $2/3$ with increasing the code distance, slightly above the threshold ($\eta\approx\eta_\text{th} + 0.01\pi$) this increasing concentration can be observed only for codes containing only even-weight stabilizers.

\begin{figure}
	\includegraphics{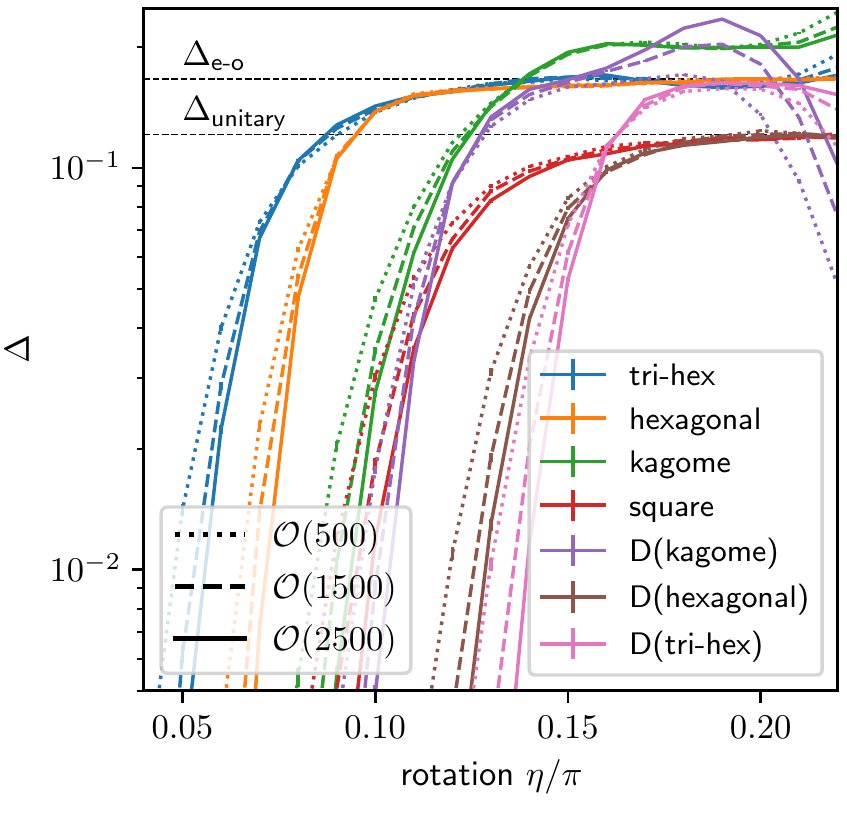}
	\caption{The syndrome average
	$\Delta$ [cf. Eq.~\eqref{eq:Delta}] for a decoder that chooses a Pauli correction optimizing $\langle\delta_s^2\rangle_{\Omega}$ for each $s$. Different colors correspond to  different $Z$-stabilizer lattices. (D denotes the dual of a lattice.) The value of $\Delta$ for strong rotations depends on the graph class; the estimate Eq.~\eqref{eq:Delta_mixed} [Eq.~\eqref{eq:Delta_unitary}] for lattices with even- and odd-weight stabilizers (even-weight stabilizer and odd-weight logical operators) is indicated by $\Delta_\text{e-o}$ ($\Delta_\text{unitary}$). The system sizes are the same as those considered in Fig.~\ref{fig:threshold}.}
	\label{fig:paulidist}
\end{figure}

\section{Noise decoherence thresholds and the coherent decoder}
\label{sec:OptDec}

The code-distance dependence of the final-state distributions in Fig.~\ref{fig:distr} suggests that, at least for certain lattices,  a second threshold $\eta_\text{th}^\text{c}$ might exist such that increasing the code distance makes the logical-level noise  increasingly Pauli like only for $\eta<\eta_\text{th}^\text{c}$. 
To study the existence of such a decoherence threshold, we first invoke a notion~\cite{bealeQuantumErrorCorrection2018} of
coherence for the operation $\mathcal{R}_s\circ\mathcal{E}$ on the logical subspace. For a logical state $\rho$, we have (with proportionality factor $|k_{s}|^{2}+|l_{s}|^{2}$)
\begin{equation}\label{eq:Pauil_coh}
 	\mathcal{R}_s\circ\mathcal{E}[\rho] \propto (1 - \epsilon^\text{P}_s) \rho + \epsilon^\text{P}_s Z_\text{L} \rho Z_\text{L} + \epsilon^\text{c}_s  Z_\text{L} \rho + \overline{\epsilon^\text{c}_s} \rho Z_\text{L} ,
\end{equation}
where $\epsilon^\text{P}_s=|l_{s}|^{2}/(|k_{s}|^{2}+|l_{s}|^{2})$ and $|\epsilon^\text{c}_s|=\sqrt{\epsilon^\text{P}_{s}(1-\epsilon^\text{P}_{s})}$. The coherent part can be defined~\cite{bealeQuantumErrorCorrection2018} as the non-Pauli contribution to Eq.~\eqref{eq:Pauil_coh}, quantified by $\epsilon^\text{c}_s$. 
The coherent part is much smaller than the Pauli part if $|\epsilon^\text{c}_s|\ll1$,  i.e., if $\epsilon^\text{P}_s$ is either close to zero or one.
(A good Pauli approximation thus requires  action that is either nearly pure identity or nearly pure $Z_\text{L}$.) This precisely corresponds to $\langle\delta_s^2\rangle_{\Omega}$ being $0$ or $2/3$ because $\langle\delta_s^2\rangle_{\Omega}=2\epsilon^\text{P}_s/3$. 

We next introduce a linearized proxy for $|\epsilon^\text{c}_s|$:
\begin{equation}\label{eq:dsc}
\langle\delta_s^2\rangle_{\Omega}^\text{c} = \frac{2}{3}\min [\epsilon^\text{P}_s,(1-\epsilon^\text{P}_s)].
\end{equation}
The less $|\epsilon^\text{c}_s|$ is, the less is $\langle\delta_s^2\rangle_{\Omega}^\text{c}$,  and vice versa. 
The distribution for $\langle\delta_s^2\rangle_{\Omega}^\text{c}$ is obtained from the $\langle\delta_s^2\rangle_{\Omega}$ distribution by mirroring, around $1/3$, the part above $1/3$ into the lower values.

The quantity $\langle\delta_s^2\rangle_{\Omega}^\text{c}$ can also be interpreted as the average infidelity obtained by choosing between the Pauli correction $C_s$ and its alternative, $C_s^\prime=C_s Z_\text{L}$ such that it minimizes $\langle\delta_s^2\rangle_{\Omega}$. With the alternative correction
\begin{equation}
\langle{\delta_s'}^2\rangle_{\Omega} = \frac{|\braket{-_\text{L}|Z_\text{L} D_s|+_\text{L}}|^2}{P(s)} = \frac{1}{3} \left( 1 + \frac{ \Re \gamma_s}{ P(s)} \right)
\end{equation}
from where we get the expression 
\begin{equation}\label{eq:cohdecode}
\begin{split}
\langle\delta_s^2\rangle_{\Omega}^\text{c} \equiv \min\left(\frac{1}{3} \left( 1 - \frac{ \Re \gamma_s}{ P(s)} \right), \frac{1}{3} \left( 1 + \frac{ \Re \gamma_s}{ P(s)} \right)\right) =  \\ =\frac{1}{3} \left( 1 - \left| \frac{ \Re \gamma_s}{ P(s)} \right| \right).
\end{split}
\end{equation}
The latter interpretation is reminiscent of the optimal decoder~\cite{dennisTopologicalQuantumMemory2002,wangSurfaceCodeQuantum2011,duclos-cianciFastDecodersTopological2010,woottonHighThresholdError2012, hutterEfficientMarkovChain2014,bravyiEfficientAlgorithmsMaximum2014}:
this calculates, given an error model, whether the syndrome $s$ is more likely to require correction with $C_s$ or $C_s Z_\text{L}$. For the optimal decoder for incoherent errors, this calculation holds for any initial logical state $\rho$ and requires no choice of error measure. 
Here, we optimize the Bloch-sphere-average and target a concrete error measure. [A $\rho$-independent variant optimizing $\theta_s$ is, however, possible for lattices corresponding to Eq.~\eqref{eq:unitarychannel}.] 
Our approach is tailored for coherent errors: it takes advantage of the deterministic nature of these in an essential manner. 
Hence, we can refer to $\langle\delta_s^2\rangle_{\Omega}^\text{c}$ as the average infidelity for the ``coherent" decoder.

To assess whether a decoherence threshold $\eta_\text{th}^\text{c}$ exists, we study
\begin{equation}\label{eq:Delta}
	\Delta = \sum_s P(s) \langle\delta_s^2\rangle_{\Omega}^\text{c},
\end{equation}
which we simulated using Monte Carlo sampling. The results obtained for the lattices and sizes introduced in Sec.~\ref{sec:threshold} are shown in Fig.~\ref{fig:paulidist}. In these averages, we can readily observe qualitative effects related to the three $Z$-stabilizer graph classes introduced in Sec.~\ref{sec:residualcoherence}: the three classes can be distinguished by the plateau value of $\Delta$ attained when the rotation parameter $\eta$ is sufficiently beyond $\eta_\text{th}$. (A different behavior sets in upon approaching the $S$-gate limit $\eta\approx \pi/4$; we henceforth focus on $\eta$ sufficiently below this value.) For $\eta$ in this regime, we find that the codes with even-weight $Z$-stabilizers and odd-weight $Z_\text{L}$ have the smallest $\Delta$, followed by those with odd-weight $Z$-stabilizers; those with even-weight $Z$-stabilizers and even-weight $Z_\text{L}$ have the largest $\Delta$.

The plateau value of $\Delta$ for lattices with odd-weight $Z$-stabilizers can be estimated by assuming that the ensemble resulting from the action of the coherent decoder on the initial state $\ket{+_\text{L}}$ corresponds to states uniformly distributed over the Bloch hemisphere closest to $\ket{+_\text{L}}$. This results in
\begin{equation}\label{eq:Delta_mixed}
	\Delta_\text{e-o} =  \frac{2}{3} \frac{1}{2 \pi} \int d\theta \sin \theta \int_{-\pi / 2}^{\pi / 2} d\phi |\braket{\theta,\phi|-_\text{L}}|^2 = \frac{1}{6}.
\end{equation} 
We can estimate the plateau of $\Delta$ for lattices with even-weight $Z$-stabilizers and odd-weight $Z_\text{L}$ similarly. Since in these lattices the accessible states, starting from $\ket{+_\text{L}}$, are the part of the equator spanning $\ket{Y_\text{L}}$ over $\ket{+_\text{L}}$ to $\ket{-Y_\text{L}}$, the average is given by
\begin{equation}\label{eq:Delta_unitary}
	\Delta_\text{unitary} = \frac{2}{3}\frac{1}{\pi} \int_{-\pi / 2}^{\pi / 2} d\phi |\braket{\theta = \frac{\pi}{2},\phi|-_\text{L}}|^2 = \frac{\pi-2}{3\pi}.
\end{equation}
Both of these values are indicated in Fig.~\ref{fig:paulidist} and fit well to the simulation.
In the case of even-weight $Z$-stabilizers and even-weight $Z_\text{L}$,  an estimate based on the final states evenly distributed among the accessible states would give the same value as Eq.~\eqref{eq:Delta_unitary}, but this is not the distribution we numerically observe. Instead, we find that the final states are increasingly concentrated around $\ket{0_\text{L}}$ and $\ket{1_\text{L}}$ upon increasing $\eta$. This explains the significantly higher plateau value in comparison to the lattices with even-weight $Z$-stabilizers and odd-weight $Z_\text{L}$. 

We now investigate the existence of a decoherence threshold $\eta_\text{th}^\text{c}$. To this end, we study the code-distance dependence of $\Delta$. As already suggested by the final-state distributions in Fig.~\ref{fig:distr}, we find qualitatively different behavior for graphs that include odd-weight $Z$-stabilizers and for those with even-weight $Z$-stabilizers. In the former case, using a similar fitting procedure as in Sec.~\ref{sec:threshold}, we observe  a decoherence threshold at $\eta_\text{th}^\text{c}\approx\eta_\text{th}+0.01\pi$: the value of $\Delta$ decreases with code distance only for $\eta<\eta_\text{th}^\text{c}$ but it increases for $\eta>\eta_\text{th}^\text{c}$. For graphs with even-weight $Z$-stabilizers, we find that if a threshold exists, it is at a much higher value of $\eta$, however, we could not clearly establish threshold behavior. 

These observations highlight that the sense in which increasing the code distance decoheres logical level noise depends on the graph class for $\eta\gtrsim \eta_\text{th}$. While for graphs with even-weight $Z$-stabilizers our findings are consistent with the final-state distribution being increasingly well approximated by that resulting from a distribution of Pauli errors, for the complementary graph class  $\eta_\text{th}^\text{c}\gtrsim \eta_\text{th}$ implies that the action of $\mathcal{R}_s \circ \mathcal{E}$ on the logical subspace can retain significant coherence for $\eta\gtrsim \eta_\text{th}$; the impact of the coherent part of the logical error is suppressed only upon averaging $\rho_s$ over syndromes (i.e., only on the average-logical-channel level).

\section{Conclusion}
We described how the $C4$-encoding of qubits can be used to obtain a  Majorana-fermion representation of  surface codes on arbitrary planar graphs, and we characterized logical-state storage under  coherent $Z$-rotations (or coherent $X$-rotations) using FLO-based simulations. 
These methods generalize the approach introduced for the square lattice by Ref.~\cite{bravyiCorrectingCoherentErrors2018}.

We studied surface codes on lattices with varying connectivity and estimated the average-logical-channel threshold values $\eta_\text{th}$ of the rotation parameter. 
Comparing  $\eta_\text{th}$ to the thresholds $\eta^\text{P}_\text{th}$ for the Pauli-twirl of the physical-qubit coherent error, we found that while $\eta_\text{th}$ and $\eta_\text{th}^\text{P}$ are similar, the inequality $\eta_\text{th}\leq \eta_\text{th}^\text{P}$ holds for all considered systems. 
We also found that, analogously to the case of incoherent Pauli noise, there is a trade-off between resilience against coherent $X$- and $Z$-rotations depending on the graph connectivity. 
However, while for Pauli noise the thresholds against bit- and phase-flips approach a universal bound, Eq.~\eqref{eq:bound}, this is not the case for incoherent errors. 
To demonstrate this, we have identified the doubly-odd lattice that is self-dual (hence has the same threshold for $Z$- and $X$-rotations) just as the square lattice, but has higher $\eta_\text{th}$ than the square lattice.

We also studied the properties of final states corresponding to individual syndrome measurements followed by recovery. 
These properties were found to follow a categorization of codes into three  classes: those whose  $Z$-stabilizers include odd-weight operators, those with only even-weight $Z$-stabilizers and even-weight logical $Z$-operator $Z_\text{L}$, and those with only even-weight $Z$-stabilizers and odd-weight $Z_\text{L}$. 
The three classes correspond to three distinct patterns of accessible final states, as shown in Fig.~\ref{fig:diffin}(b). 

The square lattice studied in Ref.~\cite{bravyiCorrectingCoherentErrors2018} corresponds to the third class; it is only in this class that per-syndrome error and recovery $\mathcal{R}_s\circ\mathcal{E}$ corresponds to a unitary $Z_\text{L}$-rotation of logical states, with state-independent syndrome probability and rotation angle. 
In all other cases, the syndrome probability depends on the initial state. To assess the average case (in the sense of this dependence), we studied the distribution of the average infidelity conditioned on syndrome $s$ [Fig.~\ref{fig:distr}], and introduced a measure of coherence [Eq.~\eqref{eq:dsc} and Fig.~\ref{fig:paulidist}] and the related coherent decoder. 
While for $\eta < \eta_\text{th}$, upon increasing the code distance the distributions are increasingly well approximated by those resulting from a distribution of Pauli errors, codes that include odd-weight $Z$-stabilizers were found to display a decoherence threshold $\eta_\text{th}^\text{c}\approx \eta_\text{th}+0.01\pi$ above which increasing code distance increases the coherence of the logical-level noise. 
The sense in which logical-level noise decoheres for $\eta\gtrsim \eta_\text{th}$ therefore depends on the graph class. In particular, for graphs with odd-weight $Z$-stabilizers, the action of $\mathcal{R}_s \circ \mathcal{E}$ on the logical subspace can retain significant coherence so that the decoherence of the logical-level noise holds only upon averaging $\rho_s$ over syndromes, i.e., only on the average-logical-channel level.

That correcting coherent errors is possible in all graph classes is an encouraging result. It shows that a unitary action for $\mathcal{R}_s\circ\mathcal{E}$ in the logical subspace, as for the square-lattice case of Ref.~\cite{bravyiCorrectingCoherentErrors2018}, is not a key requirement.  
However, our simulations are still constrained to uniaxial rotations along one of the directions specified by the stabilizers (i.e., $Z$- or $X$-rotations). It will be interesting to investigate  more general situations, including more general forms of coherent rotations, or error models with the probabilistic occurrence of different coherent components such that the overall error is inequivalent to Pauli noise.

\acknowledgments

We thank S. Brierley for bringing coherent errors, in particular Ref.~\cite{bravyiCorrectingCoherentErrors2018}, to our attention, and for a number of encouraging conversations. We acknowledge useful discussions with J. Bausch, A. Farjami, J. M. Martinis, J. K. Pachos, and S. Subramanian. This research was supported by the European Commission via the ERC Starting Grant No. 678795 TopInSy.


\begin{thebibliography}{46}%
\makeatletter
\providecommand \@ifxundefined [1]{%
 \@ifx{#1\undefined}
}%
\providecommand \@ifnum [1]{%
 \ifnum #1\expandafter \@firstoftwo
 \else \expandafter \@secondoftwo
 \fi
}%
\providecommand \@ifx [1]{%
 \ifx #1\expandafter \@firstoftwo
 \else \expandafter \@secondoftwo
 \fi
}%
\providecommand \natexlab [1]{#1}%
\providecommand \enquote  [1]{``#1''}%
\providecommand \bibnamefont  [1]{#1}%
\providecommand \bibfnamefont [1]{#1}%
\providecommand \citenamefont [1]{#1}%
\providecommand \href@noop [0]{\@secondoftwo}%
\providecommand \href [0]{\begingroup \@sanitize@url \@href}%
\providecommand \@href[1]{\@@startlink{#1}\@@href}%
\providecommand \@@href[1]{\endgroup#1\@@endlink}%
\providecommand \@sanitize@url [0]{\catcode `\\12\catcode `\$12\catcode
  `\&12\catcode `\#12\catcode `\^12\catcode `\_12\catcode `\%12\relax}%
\providecommand \@@startlink[1]{}%
\providecommand \@@endlink[0]{}%
\providecommand \url  [0]{\begingroup\@sanitize@url \@url }%
\providecommand \@url [1]{\endgroup\@href {#1}{\urlprefix }}%
\providecommand \urlprefix  [0]{URL }%
\providecommand \Eprint [0]{\href }%
\providecommand \doibase [0]{http://dx.doi.org/}%
\providecommand \selectlanguage [0]{\@gobble}%
\providecommand \bibinfo  [0]{\@secondoftwo}%
\providecommand \bibfield  [0]{\@secondoftwo}%
\providecommand \translation [1]{[#1]}%
\providecommand \BibitemOpen [0]{}%
\providecommand \bibitemStop [0]{}%
\providecommand \bibitemNoStop [0]{.\EOS\space}%
\providecommand \EOS [0]{\spacefactor3000\relax}%
\providecommand \BibitemShut  [1]{\csname bibitem#1\endcsname}%
\let\auto@bib@innerbib\@empty
%</preamble>
\bibitem [{\citenamefont {Barends}\ \emph {et~al.}(2014)\citenamefont {Barends}
  \emph {et~al.}}]{barendsSuperconductingQuantumCircuits2014a}%
  \BibitemOpen
  \bibfield  {author} {\bibinfo {author} {\bibfnamefont {R.}~\bibnamefont
  {Barends}} \emph {et~al.},\ }\href@noop {} {\bibfield  {journal} {\bibinfo
  {journal} {Nature}\ }\textbf {\bibinfo {volume} {508}},\ \bibinfo {pages}
  {500} (\bibinfo {year} {2014})}\BibitemShut {NoStop}%
\bibitem [{\citenamefont {Yan}\ \emph {et~al.}(2016)\citenamefont {Yan} \emph
  {et~al.}}]{yanFluxQubitRevisited2016}%
  \BibitemOpen
  \bibfield  {author} {\bibinfo {author} {\bibfnamefont {F.}~\bibnamefont
  {Yan}} \emph {et~al.},\ }\href@noop {} {\bibfield  {journal} {\bibinfo
  {journal} {Nat. Commun.}\ }\textbf {\bibinfo {volume} {7}},\ \bibinfo {pages}
  {12964} (\bibinfo {year} {2016})}\BibitemShut {NoStop}%
\bibitem [{\citenamefont {Kjaergaard}\ \emph {et~al.}(2020)\citenamefont
  {Kjaergaard}, \citenamefont {Schwartz}, \citenamefont {Braum{\"u}ller},
  \citenamefont {Krantz}, \citenamefont {Wang}, \citenamefont {Gustavsson},\
  and\ \citenamefont {Oliver}}]{kjaergaardSuperconductingQubitsCurrent2020}%
  \BibitemOpen
  \bibfield  {author} {\bibinfo {author} {\bibfnamefont {M.}~\bibnamefont
  {Kjaergaard}}, \bibinfo {author} {\bibfnamefont {M.~E.}\ \bibnamefont
  {Schwartz}}, \bibinfo {author} {\bibfnamefont {J.}~\bibnamefont
  {Braum{\"u}ller}}, \bibinfo {author} {\bibfnamefont {P.}~\bibnamefont
  {Krantz}}, \bibinfo {author} {\bibfnamefont {J.~I.-J.}\ \bibnamefont {Wang}},
  \bibinfo {author} {\bibfnamefont {S.}~\bibnamefont {Gustavsson}}, \ and\
  \bibinfo {author} {\bibfnamefont {W.~D.}\ \bibnamefont {Oliver}},\
  }\href@noop {} {\bibfield  {journal} {\bibinfo  {journal} {Annu. Rev.
  Condens. Matter Phys.}\ }\textbf {\bibinfo {volume} {11}},\ \bibinfo {pages}
  {369} (\bibinfo {year} {2020})}\BibitemShut {NoStop}%
\bibitem [{\citenamefont {Calderbank}\ and\ \citenamefont
  {Shor}(1996)}]{calderbankGoodQuantumErrorcorrecting1996}%
  \BibitemOpen
  \bibfield  {author} {\bibinfo {author} {\bibfnamefont {A.~R.}\ \bibnamefont
  {Calderbank}}\ and\ \bibinfo {author} {\bibfnamefont {P.~W.}\ \bibnamefont
  {Shor}},\ }\href@noop {} {\bibfield  {journal} {\bibinfo  {journal} {Phys.
  Rev. A}\ }\textbf {\bibinfo {volume} {54}},\ \bibinfo {pages} {1098}
  (\bibinfo {year} {1996})}\BibitemShut {NoStop}%
\bibitem [{\citenamefont {Steane}(1996)}]{steaneErrorCorrectingCodes1996}%
  \BibitemOpen
  \bibfield  {author} {\bibinfo {author} {\bibfnamefont {A.~M.}\ \bibnamefont
  {Steane}},\ }\href@noop {} {\bibfield  {journal} {\bibinfo  {journal} {Phys.
  Rev. Lett.}\ }\textbf {\bibinfo {volume} {77}},\ \bibinfo {pages} {793}
  (\bibinfo {year} {1996})}\BibitemShut {NoStop}%
\bibitem [{\citenamefont {Kelly}\ \emph {et~al.}(2015)\citenamefont {Kelly}
  \emph {et~al.}}]{kellyStatePreservationRepetitive2015}%
  \BibitemOpen
  \bibfield  {author} {\bibinfo {author} {\bibfnamefont {J.}~\bibnamefont
  {Kelly}} \emph {et~al.},\ }\href@noop {} {\bibfield  {journal} {\bibinfo
  {journal} {Nature}\ }\textbf {\bibinfo {volume} {519}},\ \bibinfo {pages}
  {66} (\bibinfo {year} {2015})}\BibitemShut {NoStop}%
\bibitem [{\citenamefont {Takita}\ \emph {et~al.}(2016)\citenamefont {Takita},
  \citenamefont {C{\'o}rcoles}, \citenamefont {Magesan}, \citenamefont {Abdo},
  \citenamefont {Brink}, \citenamefont {Cross}, \citenamefont {Chow},\ and\
  \citenamefont {Gambetta}}]{takitaDemonstrationWeightFourParity2016}%
  \BibitemOpen
  \bibfield  {author} {\bibinfo {author} {\bibfnamefont {M.}~\bibnamefont
  {Takita}}, \bibinfo {author} {\bibfnamefont {A.~D.}\ \bibnamefont
  {C{\'o}rcoles}}, \bibinfo {author} {\bibfnamefont {E.}~\bibnamefont
  {Magesan}}, \bibinfo {author} {\bibfnamefont {B.}~\bibnamefont {Abdo}},
  \bibinfo {author} {\bibfnamefont {M.}~\bibnamefont {Brink}}, \bibinfo
  {author} {\bibfnamefont {A.}~\bibnamefont {Cross}}, \bibinfo {author}
  {\bibfnamefont {J.~M.}\ \bibnamefont {Chow}}, \ and\ \bibinfo {author}
  {\bibfnamefont {J.~M.}\ \bibnamefont {Gambetta}},\ }\href@noop {} {\bibfield
  {journal} {\bibinfo  {journal} {Phys. Rev. Lett.}\ }\textbf {\bibinfo
  {volume} {117}},\ \bibinfo {pages} {210505} (\bibinfo {year}
  {2016})}\BibitemShut {NoStop}%
\bibitem [{\citenamefont {Terhal}(2015)}]{terhalQuantumErrorCorrection2015}%
  \BibitemOpen
  \bibfield  {author} {\bibinfo {author} {\bibfnamefont {B.~M.}\ \bibnamefont
  {Terhal}},\ }\href@noop {} {\bibfield  {journal} {\bibinfo  {journal} {Rev.
  Mod. Phys.}\ }\textbf {\bibinfo {volume} {87}},\ \bibinfo {pages} {307}
  (\bibinfo {year} {2015})}\BibitemShut {NoStop}%
\bibitem [{\citenamefont {Bravyi}\ and\ \citenamefont
  {Kitaev}(1998)}]{bravyiQuantumCodesLattice1998}%
  \BibitemOpen
  \bibfield  {author} {\bibinfo {author} {\bibfnamefont {S.~B.}\ \bibnamefont
  {Bravyi}}\ and\ \bibinfo {author} {\bibfnamefont {A.~Y.}\ \bibnamefont
  {Kitaev}},\ }\href@noop {} {\bibfield  {journal} {\bibinfo  {journal}
  {arXiv:quant-ph/9811052}\ } (\bibinfo {year} {1998})}\BibitemShut {NoStop}%
\bibitem [{\citenamefont
  {Kitaev}(2003)}]{kitaevFaulttolerantQuantumComputation2003}%
  \BibitemOpen
  \bibfield  {author} {\bibinfo {author} {\bibfnamefont {A.~Y.}\ \bibnamefont
  {Kitaev}},\ }\href@noop {} {\bibfield  {journal} {\bibinfo  {journal} {Ann.
  Phys.}\ }\textbf {\bibinfo {volume} {303}},\ \bibinfo {pages} {2} (\bibinfo
  {year} {2003})}\BibitemShut {NoStop}%
\bibitem [{\citenamefont {Fowler}\ \emph
  {et~al.}(2012{\natexlab{a}})\citenamefont {Fowler}, \citenamefont
  {Mariantoni}, \citenamefont {Martinis},\ and\ \citenamefont
  {Cleland}}]{fowlerSurfaceCodesPractical2012}%
  \BibitemOpen
  \bibfield  {author} {\bibinfo {author} {\bibfnamefont {A.~G.}\ \bibnamefont
  {Fowler}}, \bibinfo {author} {\bibfnamefont {M.}~\bibnamefont {Mariantoni}},
  \bibinfo {author} {\bibfnamefont {J.~M.}\ \bibnamefont {Martinis}}, \ and\
  \bibinfo {author} {\bibfnamefont {A.~N.}\ \bibnamefont {Cleland}},\
  }\href@noop {} {\bibfield  {journal} {\bibinfo  {journal} {Phys. Rev. A}\
  }\textbf {\bibinfo {volume} {86}} (\bibinfo {year}
  {2012}{\natexlab{a}})}\BibitemShut {NoStop}%
\bibitem [{\citenamefont {Dennis}\ \emph {et~al.}(2002)\citenamefont {Dennis},
  \citenamefont {Kitaev}, \citenamefont {Landahl},\ and\ \citenamefont
  {Preskill}}]{dennisTopologicalQuantumMemory2002}%
  \BibitemOpen
  \bibfield  {author} {\bibinfo {author} {\bibfnamefont {E.}~\bibnamefont
  {Dennis}}, \bibinfo {author} {\bibfnamefont {A.}~\bibnamefont {Kitaev}},
  \bibinfo {author} {\bibfnamefont {A.}~\bibnamefont {Landahl}}, \ and\
  \bibinfo {author} {\bibfnamefont {J.}~\bibnamefont {Preskill}},\ }\href@noop
  {} {\bibfield  {journal} {\bibinfo  {journal} {J. Math. Phys.}\ }\textbf
  {\bibinfo {volume} {43}},\ \bibinfo {pages} {4452} (\bibinfo {year}
  {2002})}\BibitemShut {NoStop}%
\bibitem [{\citenamefont {Aaronson}\ and\ \citenamefont
  {Gottesman}(2004)}]{aaronsonImprovedSimulationStabilizer2004}%
  \BibitemOpen
  \bibfield  {author} {\bibinfo {author} {\bibfnamefont {S.}~\bibnamefont
  {Aaronson}}\ and\ \bibinfo {author} {\bibfnamefont {D.}~\bibnamefont
  {Gottesman}},\ }\href@noop {} {\bibfield  {journal} {\bibinfo  {journal}
  {Phys. Rev. A}\ }\textbf {\bibinfo {volume} {70}} (\bibinfo {year}
  {2004})}\BibitemShut {NoStop}%
\bibitem [{\citenamefont {Fowler}\ \emph {et~al.}(2009)\citenamefont {Fowler},
  \citenamefont {Stephens},\ and\ \citenamefont
  {Groszkowski}}]{fowlerHighthresholdUniversalQuantum2009}%
  \BibitemOpen
  \bibfield  {author} {\bibinfo {author} {\bibfnamefont {A.~G.}\ \bibnamefont
  {Fowler}}, \bibinfo {author} {\bibfnamefont {A.~M.}\ \bibnamefont
  {Stephens}}, \ and\ \bibinfo {author} {\bibfnamefont {P.}~\bibnamefont
  {Groszkowski}},\ }\href@noop {} {\bibfield  {journal} {\bibinfo  {journal}
  {Phys. Rev. A}\ }\textbf {\bibinfo {volume} {80}},\ \bibinfo {pages} {052312}
  (\bibinfo {year} {2009})}\BibitemShut {NoStop}%
\bibitem [{\citenamefont {Wang}\ \emph {et~al.}(2011)\citenamefont {Wang},
  \citenamefont {Fowler},\ and\ \citenamefont
  {Hollenberg}}]{wangSurfaceCodeQuantum2011}%
  \BibitemOpen
  \bibfield  {author} {\bibinfo {author} {\bibfnamefont {D.~S.}\ \bibnamefont
  {Wang}}, \bibinfo {author} {\bibfnamefont {A.~G.}\ \bibnamefont {Fowler}}, \
  and\ \bibinfo {author} {\bibfnamefont {L.~C.~L.}\ \bibnamefont
  {Hollenberg}},\ }\href@noop {} {\bibfield  {journal} {\bibinfo  {journal}
  {Phys. Rev. A}\ }\textbf {\bibinfo {volume} {83}},\ \bibinfo {pages} {020302}
  (\bibinfo {year} {2011})}\BibitemShut {NoStop}%
\bibitem [{\citenamefont {Aliferis}\ \emph {et~al.}(2006)\citenamefont
  {Aliferis}, \citenamefont {Gottesman},\ and\ \citenamefont
  {Preskill}}]{aliferisQuantumAccuracyThreshold2006}%
  \BibitemOpen
  \bibfield  {author} {\bibinfo {author} {\bibfnamefont {P.}~\bibnamefont
  {Aliferis}}, \bibinfo {author} {\bibfnamefont {D.}~\bibnamefont {Gottesman}},
  \ and\ \bibinfo {author} {\bibfnamefont {J.}~\bibnamefont {Preskill}},\
  }\href@noop {} {\bibfield  {journal} {\bibinfo  {journal} {Quantum Inf.
  Comput.}\ }\textbf {\bibinfo {volume} {6}},\ \bibinfo {pages} {97} (\bibinfo
  {year} {2006})}\BibitemShut {NoStop}%
\bibitem [{\citenamefont {Chamberland}\ \emph {et~al.}(2017)\citenamefont
  {Chamberland}, \citenamefont {Wallman}, \citenamefont {Beale},\ and\
  \citenamefont {Laflamme}}]{chamberlandHardDecodingAlgorithm2017}%
  \BibitemOpen
  \bibfield  {author} {\bibinfo {author} {\bibfnamefont {C.}~\bibnamefont
  {Chamberland}}, \bibinfo {author} {\bibfnamefont {J.}~\bibnamefont
  {Wallman}}, \bibinfo {author} {\bibfnamefont {S.}~\bibnamefont {Beale}}, \
  and\ \bibinfo {author} {\bibfnamefont {R.}~\bibnamefont {Laflamme}},\
  }\href@noop {} {\bibfield  {journal} {\bibinfo  {journal} {Phys. Rev. A}\
  }\textbf {\bibinfo {volume} {95}},\ \bibinfo {pages} {042332} (\bibinfo
  {year} {2017})}\BibitemShut {NoStop}%
\bibitem [{\citenamefont {Guti{\'e}rrez}\ \emph {et~al.}(2016)\citenamefont
  {Guti{\'e}rrez}, \citenamefont {Smith}, \citenamefont {Lulushi},
  \citenamefont {Janardan},\ and\ \citenamefont
  {Brown}}]{gutierrezErrorsPseudothresholdsIncoherent2016}%
  \BibitemOpen
  \bibfield  {author} {\bibinfo {author} {\bibfnamefont {M.}~\bibnamefont
  {Guti{\'e}rrez}}, \bibinfo {author} {\bibfnamefont {C.}~\bibnamefont
  {Smith}}, \bibinfo {author} {\bibfnamefont {L.}~\bibnamefont {Lulushi}},
  \bibinfo {author} {\bibfnamefont {S.}~\bibnamefont {Janardan}}, \ and\
  \bibinfo {author} {\bibfnamefont {K.~R.}\ \bibnamefont {Brown}},\ }\href@noop
  {} {\bibfield  {journal} {\bibinfo  {journal} {Phys. Rev. A}\ }\textbf
  {\bibinfo {volume} {94}},\ \bibinfo {pages} {042338} (\bibinfo {year}
  {2016})}\BibitemShut {NoStop}%
\bibitem [{\citenamefont {Cai}\ \emph {et~al.}(2020)\citenamefont {Cai},
  \citenamefont {Xu},\ and\ \citenamefont
  {Benjamin}}]{caiMitigatingCoherentNoise2020}%
  \BibitemOpen
  \bibfield  {author} {\bibinfo {author} {\bibfnamefont {Z.}~\bibnamefont
  {Cai}}, \bibinfo {author} {\bibfnamefont {X.}~\bibnamefont {Xu}}, \ and\
  \bibinfo {author} {\bibfnamefont {S.~C.}\ \bibnamefont {Benjamin}},\
  }\href@noop {} {\bibfield  {journal} {\bibinfo  {journal} {npj Quantum Inf.}\
  }\textbf {\bibinfo {volume} {6}},\ \bibinfo {pages} {17} (\bibinfo {year}
  {2020})}\BibitemShut {NoStop}%
\bibitem [{\citenamefont
  {Gottesman}(2019)}]{gottesmanMaximallySensitiveSets2019}%
  \BibitemOpen
  \bibfield  {author} {\bibinfo {author} {\bibfnamefont {D.}~\bibnamefont
  {Gottesman}},\ }\href@noop {} {\bibfield  {journal} {\bibinfo  {journal}
  {arXiv:1907.05950 [quant-ph]}\ } (\bibinfo {year} {2019})}\BibitemShut
  {NoStop}%
\bibitem [{\citenamefont {Sanders}\ \emph {et~al.}(2015)\citenamefont
  {Sanders}, \citenamefont {Wallman},\ and\ \citenamefont
  {Sanders}}]{sandersBoundingQuantumGate2015}%
  \BibitemOpen
  \bibfield  {author} {\bibinfo {author} {\bibfnamefont {Y.~R.}\ \bibnamefont
  {Sanders}}, \bibinfo {author} {\bibfnamefont {J.~J.}\ \bibnamefont
  {Wallman}}, \ and\ \bibinfo {author} {\bibfnamefont {B.~C.}\ \bibnamefont
  {Sanders}},\ }\href@noop {} {\bibfield  {journal} {\bibinfo  {journal} {New
  J. Phys.}\ }\textbf {\bibinfo {volume} {18}},\ \bibinfo {pages} {012002}
  (\bibinfo {year} {2015})}\BibitemShut {NoStop}%
\bibitem [{\citenamefont {Huang}\ \emph {et~al.}(2019)\citenamefont {Huang},
  \citenamefont {Doherty},\ and\ \citenamefont
  {Flammia}}]{huangPerformanceQuantumError2019}%
  \BibitemOpen
  \bibfield  {author} {\bibinfo {author} {\bibfnamefont {E.}~\bibnamefont
  {Huang}}, \bibinfo {author} {\bibfnamefont {A.~C.}\ \bibnamefont {Doherty}},
  \ and\ \bibinfo {author} {\bibfnamefont {S.}~\bibnamefont {Flammia}},\
  }\href@noop {} {\bibfield  {journal} {\bibinfo  {journal} {Phys. Rev. A}\
  }\textbf {\bibinfo {volume} {99}},\ \bibinfo {pages} {022313} (\bibinfo
  {year} {2019})}\BibitemShut {NoStop}%
\bibitem [{\citenamefont {Greenbaum}\ and\ \citenamefont
  {Dutton}(2018)}]{greenbaumModelingCoherentErrors2018}%
  \BibitemOpen
  \bibfield  {author} {\bibinfo {author} {\bibfnamefont {D.}~\bibnamefont
  {Greenbaum}}\ and\ \bibinfo {author} {\bibfnamefont {Z.}~\bibnamefont
  {Dutton}},\ }\href@noop {} {\bibfield  {journal} {\bibinfo  {journal}
  {Quantum Sci. Technol.}\ }\textbf {\bibinfo {volume} {3}},\ \bibinfo {pages}
  {015007} (\bibinfo {year} {2018})}\BibitemShut {NoStop}%
\bibitem [{\citenamefont {Bravyi}\ \emph {et~al.}(2018)\citenamefont {Bravyi},
  \citenamefont {Englbrecht}, \citenamefont {K{\"o}nig},\ and\ \citenamefont
  {Peard}}]{bravyiCorrectingCoherentErrors2018}%
  \BibitemOpen
  \bibfield  {author} {\bibinfo {author} {\bibfnamefont {S.}~\bibnamefont
  {Bravyi}}, \bibinfo {author} {\bibfnamefont {M.}~\bibnamefont {Englbrecht}},
  \bibinfo {author} {\bibfnamefont {R.}~\bibnamefont {K{\"o}nig}}, \ and\
  \bibinfo {author} {\bibfnamefont {N.}~\bibnamefont {Peard}},\ }\href@noop {}
  {\bibfield  {journal} {\bibinfo  {journal} {npj Quantum Inf.}\ }\textbf
  {\bibinfo {volume} {4}},\ \bibinfo {pages} {55} (\bibinfo {year}
  {2018})}\BibitemShut {NoStop}%
\bibitem [{\citenamefont {Beale}\ \emph {et~al.}(2018)\citenamefont {Beale},
  \citenamefont {Wallman}, \citenamefont {Guti{\'e}rrez}, \citenamefont
  {Brown},\ and\ \citenamefont {Laflamme}}]{bealeQuantumErrorCorrection2018}%
  \BibitemOpen
  \bibfield  {author} {\bibinfo {author} {\bibfnamefont {S.~J.}\ \bibnamefont
  {Beale}}, \bibinfo {author} {\bibfnamefont {J.~J.}\ \bibnamefont {Wallman}},
  \bibinfo {author} {\bibfnamefont {M.}~\bibnamefont {Guti{\'e}rrez}}, \bibinfo
  {author} {\bibfnamefont {K.~R.}\ \bibnamefont {Brown}}, \ and\ \bibinfo
  {author} {\bibfnamefont {R.}~\bibnamefont {Laflamme}},\ }\href@noop {}
  {\bibfield  {journal} {\bibinfo  {journal} {Phys. Rev. Lett.}\ }\textbf
  {\bibinfo {volume} {121}},\ \bibinfo {pages} {190501} (\bibinfo {year}
  {2018})}\BibitemShut {NoStop}%
\bibitem [{\citenamefont {Iverson}\ and\ \citenamefont
  {Preskill}(2020)}]{iversonCoherenceLogicalQuantum2020}%
  \BibitemOpen
  \bibfield  {author} {\bibinfo {author} {\bibfnamefont {J.~K.}\ \bibnamefont
  {Iverson}}\ and\ \bibinfo {author} {\bibfnamefont {J.}~\bibnamefont
  {Preskill}},\ }\href@noop {} {\bibfield  {journal} {\bibinfo  {journal} {New
  J. Phys.}\ } (\bibinfo {year} {2020})}\BibitemShut {NoStop}%
\bibitem [{\citenamefont {Darmawan}\ and\ \citenamefont
  {Poulin}(2017)}]{darmawanTensorNetworkSimulationsSurface2017}%
  \BibitemOpen
  \bibfield  {author} {\bibinfo {author} {\bibfnamefont {A.~S.}\ \bibnamefont
  {Darmawan}}\ and\ \bibinfo {author} {\bibfnamefont {D.}~\bibnamefont
  {Poulin}},\ }\href@noop {} {\bibfield  {journal} {\bibinfo  {journal} {Phys.
  Rev. Lett.}\ }\textbf {\bibinfo {volume} {119}},\ \bibinfo {pages} {040502}
  (\bibinfo {year} {2017})}\BibitemShut {NoStop}%
\bibitem [{\citenamefont {Kitaev}(2006)}]{kitaevAnyonsExactlySolved2006}%
  \BibitemOpen
  \bibfield  {author} {\bibinfo {author} {\bibfnamefont {A.}~\bibnamefont
  {Kitaev}},\ }\href@noop {} {\bibfield  {journal} {\bibinfo  {journal} {Ann.
  Phys.}\ }\textbf {\bibinfo {volume} {321}},\ \bibinfo {pages} {2} (\bibinfo
  {year} {2006})}\BibitemShut {NoStop}%
\bibitem [{\citenamefont {Wen}(2003)}]{wenQuantumOrdersExact2003a}%
  \BibitemOpen
  \bibfield  {author} {\bibinfo {author} {\bibfnamefont {X.-G.}\ \bibnamefont
  {Wen}},\ }\href@noop {} {\bibfield  {journal} {\bibinfo  {journal} {Phys.
  Rev. Lett.}\ }\textbf {\bibinfo {volume} {90}},\ \bibinfo {pages} {016803}
  (\bibinfo {year} {2003})}\BibitemShut {NoStop}%
\bibitem [{\citenamefont {Terhal}\ and\ \citenamefont
  {DiVincenzo}(2002)}]{terhalClassicalSimulationNoninteractingfermion2002}%
  \BibitemOpen
  \bibfield  {author} {\bibinfo {author} {\bibfnamefont {B.~M.}\ \bibnamefont
  {Terhal}}\ and\ \bibinfo {author} {\bibfnamefont {D.~P.}\ \bibnamefont
  {DiVincenzo}},\ }\href@noop {} {\bibfield  {journal} {\bibinfo  {journal}
  {Phys. Rev. A}\ }\textbf {\bibinfo {volume} {65}},\ \bibinfo {pages} {032325}
  (\bibinfo {year} {2002})}\BibitemShut {NoStop}%
\bibitem [{\citenamefont {Fujii}\ and\ \citenamefont
  {Tokunaga}(2012)}]{fujiiErrorLossTolerances2012}%
  \BibitemOpen
  \bibfield  {author} {\bibinfo {author} {\bibfnamefont {K.}~\bibnamefont
  {Fujii}}\ and\ \bibinfo {author} {\bibfnamefont {Y.}~\bibnamefont
  {Tokunaga}},\ }\href@noop {} {\bibfield  {journal} {\bibinfo  {journal}
  {Phys. Rev. A}\ }\textbf {\bibinfo {volume} {86}} (\bibinfo {year}
  {2012})}\BibitemShut {NoStop}%
\bibitem [{\citenamefont {R{\"o}thlisberger}\ \emph {et~al.}(2012)\citenamefont
  {R{\"o}thlisberger}, \citenamefont {Wootton}, \citenamefont {Heath},
  \citenamefont {Pachos},\ and\ \citenamefont
  {Loss}}]{rothlisbergerIncoherentDynamicsToric2012}%
  \BibitemOpen
  \bibfield  {author} {\bibinfo {author} {\bibfnamefont {B.}~\bibnamefont
  {R{\"o}thlisberger}}, \bibinfo {author} {\bibfnamefont {J.~R.}\ \bibnamefont
  {Wootton}}, \bibinfo {author} {\bibfnamefont {R.~M.}\ \bibnamefont {Heath}},
  \bibinfo {author} {\bibfnamefont {J.~K.}\ \bibnamefont {Pachos}}, \ and\
  \bibinfo {author} {\bibfnamefont {D.}~\bibnamefont {Loss}},\ }\href@noop {}
  {\bibfield  {journal} {\bibinfo  {journal} {Phys. Rev. A}\ }\textbf {\bibinfo
  {volume} {85}},\ \bibinfo {pages} {022313} (\bibinfo {year}
  {2012})}\BibitemShut {NoStop}%
\bibitem [{\citenamefont
  {Gottesman}(1997)}]{gottesmanStabilizerCodesQuantum1997}%
  \BibitemOpen
  \bibfield  {author} {\bibinfo {author} {\bibfnamefont {D.}~\bibnamefont
  {Gottesman}},\ }\href@noop {} {\bibfield  {journal} {\bibinfo  {journal}
  {arXiv:quant-ph/9705052}\ } (\bibinfo {year} {1997})}\BibitemShut {NoStop}%
\bibitem [{\citenamefont {Nielsen}\ and\ \citenamefont
  {Chuang}(2000)}]{nielsenQuantumComputationQuantum2000}%
  \BibitemOpen
  \bibfield  {author} {\bibinfo {author} {\bibfnamefont {M.}~\bibnamefont
  {Nielsen}}\ and\ \bibinfo {author} {\bibfnamefont {I.}~\bibnamefont
  {Chuang}},\ }\href@noop {} {\emph {\bibinfo {title} {Quantum {{Computation}}
  and {{Quantum Information}}}}}\ (\bibinfo  {publisher} {{Cambridge University
  Press}},\ \bibinfo {address} {{Cambridge}},\ \bibinfo {year}
  {2000})\BibitemShut {NoStop}%
\bibitem [{\citenamefont
  {Kitaev}(1997)}]{kitaevQuantumComputationsAlgorithms1997}%
  \BibitemOpen
  \bibfield  {author} {\bibinfo {author} {\bibfnamefont {A.~Y.}\ \bibnamefont
  {Kitaev}},\ }\href@noop {} {\bibfield  {journal} {\bibinfo  {journal} {Russ.
  Math. Surv.}\ }\textbf {\bibinfo {volume} {52}},\ \bibinfo {pages} {1191}
  (\bibinfo {year} {1997})}\BibitemShut {NoStop}%
\bibitem [{\citenamefont {Fowler}\ \emph
  {et~al.}(2012{\natexlab{b}})\citenamefont {Fowler}, \citenamefont
  {Whiteside},\ and\ \citenamefont
  {Hollenberg}}]{fowlerPracticalClassicalProcessing2012}%
  \BibitemOpen
  \bibfield  {author} {\bibinfo {author} {\bibfnamefont {A.~G.}\ \bibnamefont
  {Fowler}}, \bibinfo {author} {\bibfnamefont {A.~C.}\ \bibnamefont
  {Whiteside}}, \ and\ \bibinfo {author} {\bibfnamefont {L.~C.~L.}\
  \bibnamefont {Hollenberg}},\ }\href@noop {} {\bibfield  {journal} {\bibinfo
  {journal} {Phys. Rev. Lett.}\ }\textbf {\bibinfo {volume} {108}} (\bibinfo
  {year} {2012}{\natexlab{b}})}\BibitemShut {NoStop}%
\bibitem [{\citenamefont {Edmonds}(1965)}]{edmondsPathsTreesFlowers1965}%
  \BibitemOpen
  \bibfield  {author} {\bibinfo {author} {\bibfnamefont {J.}~\bibnamefont
  {Edmonds}},\ }\href@noop {} {\bibfield  {journal} {\bibinfo  {journal} {Can.
  J. Math.}\ }\textbf {\bibinfo {volume} {17}},\ \bibinfo {pages} {449}
  (\bibinfo {year} {1965})}\BibitemShut {NoStop}%
\bibitem [{\citenamefont {Rahn}\ \emph {et~al.}(2002)\citenamefont {Rahn},
  \citenamefont {Doherty},\ and\ \citenamefont
  {Mabuchi}}]{rahnExactPerformanceConcatenated2002}%
  \BibitemOpen
  \bibfield  {author} {\bibinfo {author} {\bibfnamefont {B.}~\bibnamefont
  {Rahn}}, \bibinfo {author} {\bibfnamefont {A.~C.}\ \bibnamefont {Doherty}}, \
  and\ \bibinfo {author} {\bibfnamefont {H.}~\bibnamefont {Mabuchi}},\
  }\href@noop {} {\bibfield  {journal} {\bibinfo  {journal} {Phys. Rev. A}\
  }\textbf {\bibinfo {volume} {66}},\ \bibinfo {pages} {032304} (\bibinfo
  {year} {2002})}\BibitemShut {NoStop}%
\bibitem [{\citenamefont {Beigi}\ and\ \citenamefont
  {K{\"o}nig}(2011)}]{beigiSimplifiedInstantaneousNonlocal2011}%
  \BibitemOpen
  \bibfield  {author} {\bibinfo {author} {\bibfnamefont {S.}~\bibnamefont
  {Beigi}}\ and\ \bibinfo {author} {\bibfnamefont {R.}~\bibnamefont
  {K{\"o}nig}},\ }\href@noop {} {\bibfield  {journal} {\bibinfo  {journal} {New
  J. Phys.}\ }\textbf {\bibinfo {volume} {13}},\ \bibinfo {pages} {093036}
  (\bibinfo {year} {2011})}\BibitemShut {NoStop}%
\bibitem [{\citenamefont {Wallman}\ and\ \citenamefont
  {Flammia}(2014)}]{wallmanRandomizedBenchmarkingConfidence2014}%
  \BibitemOpen
  \bibfield  {author} {\bibinfo {author} {\bibfnamefont {J.~J.}\ \bibnamefont
  {Wallman}}\ and\ \bibinfo {author} {\bibfnamefont {S.~T.}\ \bibnamefont
  {Flammia}},\ }\href@noop {} {\bibfield  {journal} {\bibinfo  {journal} {New
  J. Phys.}\ }\textbf {\bibinfo {volume} {16}},\ \bibinfo {pages} {103032}
  (\bibinfo {year} {2014})}\BibitemShut {NoStop}%
\bibitem [{\citenamefont
  {Kasteleyn}(1967)}]{kasteleynGraphTheoryTheoretical1967}%
  \BibitemOpen
  \bibfield  {author} {\bibinfo {author} {\bibfnamefont {P.~W.}\ \bibnamefont
  {Kasteleyn}},\ }in\ \href@noop {} {\emph {\bibinfo {booktitle} {Graph Theory
  and Crystal Physics}}},\ \bibinfo {editor} {edited by\ \bibinfo {editor}
  {\bibfnamefont {F.}~\bibnamefont {Harary}}}\ (\bibinfo  {publisher}
  {{AcademicPress}},\ \bibinfo {address} {{New York}},\ \bibinfo {year}
  {1967})\ pp.\ \bibinfo {pages} {43--110}\BibitemShut {NoStop}%
\bibitem [{\citenamefont {Johnston}\ \emph {et~al.}(2009)\citenamefont
  {Johnston}, \citenamefont {Kribs},\ and\ \citenamefont
  {Paulsen}}]{johnstonComputingStabilizedNorms2009}%
  \BibitemOpen
  \bibfield  {author} {\bibinfo {author} {\bibfnamefont {N.}~\bibnamefont
  {Johnston}}, \bibinfo {author} {\bibfnamefont {D.~W.}\ \bibnamefont {Kribs}},
  \ and\ \bibinfo {author} {\bibfnamefont {V.~I.}\ \bibnamefont {Paulsen}},\
  }\href@noop {} {\bibfield  {journal} {\bibinfo  {journal} {Quant. Inf.
  Comput.}\ }\textbf {\bibinfo {volume} {9}},\ \bibinfo {pages} {16} (\bibinfo
  {year} {2009})}\BibitemShut {NoStop}%
\bibitem [{\citenamefont {{Duclos-Cianci}}\ and\ \citenamefont
  {Poulin}(2010)}]{duclos-cianciFastDecodersTopological2010}%
  \BibitemOpen
  \bibfield  {author} {\bibinfo {author} {\bibfnamefont {G.}~\bibnamefont
  {{Duclos-Cianci}}}\ and\ \bibinfo {author} {\bibfnamefont {D.}~\bibnamefont
  {Poulin}},\ }\href@noop {} {\bibfield  {journal} {\bibinfo  {journal} {Phys.
  Rev. Lett.}\ }\textbf {\bibinfo {volume} {104}},\ \bibinfo {pages} {050504}
  (\bibinfo {year} {2010})}\BibitemShut {NoStop}%
\bibitem [{\citenamefont {Wootton}\ and\ \citenamefont
  {Loss}(2012)}]{woottonHighThresholdError2012}%
  \BibitemOpen
  \bibfield  {author} {\bibinfo {author} {\bibfnamefont {J.~R.}\ \bibnamefont
  {Wootton}}\ and\ \bibinfo {author} {\bibfnamefont {D.}~\bibnamefont {Loss}},\
  }\href@noop {} {\bibfield  {journal} {\bibinfo  {journal} {Phys. Rev. Lett.}\
  }\textbf {\bibinfo {volume} {109}},\ \bibinfo {pages} {160503} (\bibinfo
  {year} {2012})}\BibitemShut {NoStop}%
\bibitem [{\citenamefont {Hutter}\ \emph {et~al.}(2014)\citenamefont {Hutter},
  \citenamefont {Wootton},\ and\ \citenamefont
  {Loss}}]{hutterEfficientMarkovChain2014}%
  \BibitemOpen
  \bibfield  {author} {\bibinfo {author} {\bibfnamefont {A.}~\bibnamefont
  {Hutter}}, \bibinfo {author} {\bibfnamefont {J.~R.}\ \bibnamefont {Wootton}},
  \ and\ \bibinfo {author} {\bibfnamefont {D.}~\bibnamefont {Loss}},\
  }\href@noop {} {\bibfield  {journal} {\bibinfo  {journal} {Phys. Rev. A}\
  }\textbf {\bibinfo {volume} {89}},\ \bibinfo {pages} {022326} (\bibinfo
  {year} {2014})}\BibitemShut {NoStop}%
\bibitem [{\citenamefont {Bravyi}\ \emph {et~al.}(2014)\citenamefont {Bravyi},
  \citenamefont {Suchara},\ and\ \citenamefont
  {Vargo}}]{bravyiEfficientAlgorithmsMaximum2014}%
  \BibitemOpen
  \bibfield  {author} {\bibinfo {author} {\bibfnamefont {S.}~\bibnamefont
  {Bravyi}}, \bibinfo {author} {\bibfnamefont {M.}~\bibnamefont {Suchara}}, \
  and\ \bibinfo {author} {\bibfnamefont {A.}~\bibnamefont {Vargo}},\
  }\href@noop {} {\bibfield  {journal} {\bibinfo  {journal} {Phys. Rev. A}\
  }\textbf {\bibinfo {volume} {90}},\ \bibinfo {pages} {032326} (\bibinfo
  {year} {2014})}\BibitemShut {NoStop}%
\end{thebibliography}
\end{document}